\RequirePackage[2024-06-01]{latexrelease}
\documentclass[%
 reprint,
superscriptaddress,
amsmath,
amssymb,
aps,
floatfix,
]{revtex4-2}

\usepackage{graphicx}
\usepackage{dcolumn}
\usepackage{bm}
\usepackage[dvipsnames,table]{xcolor}
\usepackage[normalem]{ulem}
\usepackage[T1]{fontenc}
\usepackage{braket}
\usepackage{siunitx}
\usepackage{tabularx}

\graphicspath{{./figures/}}

\newcommand{\CB}[1]{\textcolor{black}{#1}}

\begin{document}

\title{Quantum communication networks with defects in silicon carbide} 

\author{Philipp Sohr}
\email{philipp.sohr@student.tuwien.ac.at}
\affiliation{Institute for Quantum Optics and Quantum Information (IQOQI), Austrian Academy of Sciences, Boltzmanngasse 3, 1090 Vienna, Austria}
\affiliation{Vienna Center for Quantum Science and Technology, Atominstitut, TU Wien, 1020 Vienna, Austria}
\affiliation{Quantum Technology Laboratories GmbH (qtlabs), Clemens-Holzmeister-Straße 6/6, 1100 Vienna, Austria}
\author{Philipp Koller}
\affiliation{Institute for Quantum Optics and Quantum Information (IQOQI), Austrian Academy of Sciences, Boltzmanngasse 3, 1090 Vienna, Austria}
\affiliation{University of Vienna, Faculty of Physics \& Vienna Doctoral School in Physics,  Boltzmanngasse 5, A-1090 Vienna, Austria}
\author{Sebastian Ecker}
\affiliation{Quantum Technology Laboratories GmbH (qtlabs), Clemens-Holzmeister-Straße 6/6, 1100 Vienna, Austria}
\author{Matthias Fink}
\affiliation{Quantum Technology Laboratories GmbH (qtlabs), Clemens-Holzmeister-Straße 6/6, 1100 Vienna, Austria}
\author{Thomas Scheidl}
\affiliation{Quantum Technology Laboratories GmbH (qtlabs), Clemens-Holzmeister-Straße 6/6, 1100 Vienna, Austria}
\author{Rupert Ursin}
\affiliation{Quantum Technology Laboratories GmbH (qtlabs), Clemens-Holzmeister-Straße 6/6, 1100 Vienna, Austria}
\author{Muhammad Junaid Arshad}
\affiliation{Institute of Photonics and Quantum Sciences, SUPA, Heriot-Watt University, Edinburgh EH14 4AS, United Kingdom}
\author{Cristian Bonato}
\affiliation{Institute of Photonics and Quantum Sciences, SUPA, Heriot-Watt University, Edinburgh EH14 4AS, United Kingdom}
\author{Pasquale Cilibrizzi}
\affiliation{Institute of Photonics and Quantum Sciences, SUPA, Heriot-Watt University, Edinburgh EH14 4AS, United Kingdom}
\author{Adam Gali}
\affiliation{HUN-REN Wigner Research Centre for Physics, PO.\ Box 49, H-1525 Budapest, Hungary}
\affiliation{Department of Atomic Physics, Institute of Physics, Budapest University of Technology and Economics, M\H{u}egyetem rakpart 3., H-1111 Budapest, Hungary}
\affiliation{MTA-WFK Lend\"ulet "Momentum" Semiconductor Nanostructures Research Group, PO.\ Box 49, H-1525 Budapest, Hungary}
\author{Péter Udvarhelyi}
\affiliation{HUN-REN Wigner Research Centre for Physics, PO.\ Box 49, H-1525 Budapest, Hungary}
\affiliation{Department of Atomic Physics, Institute of Physics, Budapest University of Technology and Economics, M\H{u}egyetem rakpart 3., H-1111 Budapest, Hungary}
\author{Alberto Politi}
\author{Oliver J. Trojak }
\affiliation{School of Physics and Astronomy, University of Southampton, Southampton SO17 1BJ, United Kingdom}
\author{Misagh Ghezellou}
\author{Jawad Ul Hassan}
\author{Ivan G. Ivanov}
\author{Nguyen Tien Son}
\affiliation{University, Department of Physics, Chemistry and Biology, SE-58183 Linköping, Sweden}
\author{Guido Burkard}
\author{Benedikt Tissot}
\affiliation{Department of Physics, University of Konstanz, D-78457 Konstanz, Germany}
\author{Joop Hendriks}
\author{Carmem M. Gilardoni}
\author{Caspar H. van der Wal}
\affiliation{Zernike Institute for Advanced Materials, University of Groningen, NL-9747AG  Groningen, The Netherlands}
\author{Christian David}
\author{Masa Mokhtarzadeh}
\affiliation{Paul Scherrer Institut, CH-5232 Villigen, Switzerland}
\author{Thomas Astner}
\affiliation{Institute for Quantum Optics and Quantum Information (IQOQI), Austrian Academy of Sciences, Boltzmanngasse 3, 1090 Vienna, Austria}
\author{Michael Trupke}
\email{michael.trupke@oeaw.ac.at}
\affiliation{Institute for Quantum Optics and Quantum Information (IQOQI), Austrian Academy of Sciences, Boltzmanngasse 3, 1090 Vienna, Austria}

\begin{abstract}
Quantum communication promises unprecedented capabilities enabled by the transmission of quantum states of light.
However, current implementations face severe distance limitations due to photon loss.
Silicon carbide (SiC) defects have emerged as a promising quantum device platform, offering strong optical transitions, long spin coherence lifetimes and the opportunity for integration with semiconductor devices.
Some defects with optical transitions in the telecom range have been identified, allowing to interface with fiber networks without the need for wavelength conversion.
These unique properties make SiC an attractive platform for the implementation of quantum nodes for quantum communication networks.
We provide an overview of the most prominent defects in SiC and their implementation in spin-photon interfaces.
Furthermore, we model an exemplary, memory-enhanced quantum communication protocol in order to extract the parameters required to surpass a direct point-to-point link performance.
Based on these insights, we summarize the key steps required towards the deployment of SiC devices in large-scale quantum communication networks.

\end{abstract}

\maketitle

\section{Introduction}
The advent of the information age is driven not by computers alone, but increasingly by the connections between many computers into ever-growing information processing networks. Similarly, quantum links will prove essential in distributed quantum computing and quantum sensing~\cite{Wehner2018, Zhang2021}. A key step on this path is the exploitation of quantum properties for the communication process itself.

The exchange of photons in quantum communication enables communication primitives that are unachievable in the classical realm. 
In quantum cryptography, for example, quantum states of light are used to exchange cryptographic keys which are inaccessible by a third party by virtue of the no-cloning theorem~\cite{park1970concept,wootters1982single}.
The main roadblock towards widespread deployment of quantum cryptography is the limited communication distance of a few 100 km in optical fiber due to the absence of a quantum amplifier. 
Currently, there are two approaches of extending the transmission distance of flying qubits.
The first one involves free-space links between optical ground receivers and satellites, an approach which has seen the first in-field demonstrations in recent years~\cite{Lu2022}. 
Satellite-based quantum communication enables exchange of cryptographic keys around the globe~\cite{Liao2018,10.1117/12.2689972}, where the loss is dominated by diffraction instead of absorption. 
The second approach relies on intermediate nodes in fiber-based networks, which promise to overcome the fundamental limit of point-to-point connections~\cite{Briegel1998}.      
These nodes interface flying qubits with stationary qubits.
Not only do stationary qubits act as quantum memories to extend the transmission distance of quantum states, but they also facilitate processing and routing of quantum information in complex network topologies. 

Quantum nodes should therefore simultaneously fulfill several requirements. 
First and foremost, the transfer of quantum information to and from the nodes should be faithful and efficient~\cite{divincenzo2000physical}. 
Secondly, the storage time, which is limited by decoherence effects of the underlying physical system is crucial.
In particular, it should be long enough to enable local quantum processing in addition to quantum and classical communication, both of which are limited by the travel time of the photons over global distances and the success probability of photon transmission. 
Ideally, the state can be transferred from the spin memory to a long-lived quantum register with a storage time which is longer by orders of magnitude~\cite{Lukin2020}. 
Crucially, the memories should operate at one of the telecommunication wavelength bands in order to minimize transmission losses and guarantee compatibility with existing fiber infrastructure. 
While quantum frequency conversion is a viable solution for shifting frequencies to the telecommunication band~\cite{Dreau_Hanson_2018, Tchebotareva2019, Rakonjac_ICFO_2021}, added noise and losses are inevitable.
In addition, the nodes should be hosted on a scalable and integrated platform for parallelization and multiplexing purposes.

While several candidate systems, such as atomic ensembles~\cite{Hammerer2010}, quantum dots~\cite{quantumdot2010,Lodahl2017}, rare-earth-doped solids~\cite{Lago2021,Liu2021} and trapped ions~\cite{Krutyanskiy2023} are currently being investigated, defects in semiconductor materials are a particularly promising platform: They offer a view to scalable fabrication, integration with electronics, and compatibility with the highly advanced tools and methods of the semiconductor industry. 
Among these, a number of defects in silicon carbide have gained attention due to their strong optical transitions and long spin coherence lifetimes~\cite{Lukin2020, Castelletto2022,zhou2025SiCNetworks}.

A central concept in the deployment of semiconductor defects for quantum networks is spin-photon entanglement. 
Spin centers can be interfaced with photonic time-bin qubits as well as polarization qubits, making them versatile and deployable for both free-space and fiber-optical quantum communication~\cite{Christle2017, Vasconcelos2020, bader2024analysis, he2026spinPhoton}. These hybrid entangled states between a stationary and a flying qubit facilitate heralded entanglement between two quantum memories via Bell state measurements (BSM). 
To this end, two photons are interfered on a beamsplitter in a synchronous BSM, or a single defect is read out after two consecutive writing processes, which constitutes an asynchronous BSM~\cite{Panayi2014,Bhaskar2020}. 
BSMs are also key in entanglement swapping, which in turn is an essential building block of the quantum repeater protocol~\cite{Briegel1998}. 
While impressive progress has been achieved, including demonstrations of heralded entanglement distribution between quantum memories and basic quantum network operations~\cite{Lago2021,Liu2021,Pompili2021,bersin2023telecom, lu2026DI-QKD, liu2026long}, fully-fledged quantum repeater chains which outperform point-to-point connections in quantum cryptography remain an outstanding challenge. 
However, several quantum cryptography protocols exhibit a substantial benefit over the direct transmission of photons with only a single intermediate network node. 
One of these protocols is memory-assisted quantum key distribution (MA-QKD)~\cite{Panayi2014}, which is regarded as an important milestone on the path to outperforming direct transmission~\cite{Azuma2022}. 

In this work, we present the results of MA-QKD simulations with parameters from state-of-the-art SiC devices. 
This approach serves as a testbed for the performance of SiC memories in a relevant quantum communication setting. 
The results of this study provide the basis for the development of a roadmap for SiC devices. 
Furthermore, we discuss advanced quantum communication scenarios which benefit from the SiC platform.     
This paper is organized as follows. 
In Sec.~\ref{sec:sicdefects} we introduce the silicon carbide platform and some of the defects which are of interest,
while Sec.~\ref{sec:sicdevices} is devoted to the photonic properties of the defects and their integration in quantum devices.
We describe how these quantum devices based on SiC are suited as quantum nodes in Sec.~\ref{sec:qunode}.
After a short introduction to quantum cryptography, Sec.~\ref{sec:MAQKD} is dedicated to a particular memory-assisted QKD protocol, with secure key rate simulation results presented in Sec.~\ref{sec:results}. 
We draw a roadmap for necessary improvements of SiC devices for quantum link deployment in Sec.~\ref{sec:roadmap} and conclude with a summary in Sec.~\ref{sec:summary}.

\section{The silicon carbide platform}

\subsection{Implementing spin-photon interfaces with defects in SiC} \label{sec:sicdefects}

Single electron spins may be optically accessed through point defects in semiconductors, such as SiC~\cite{Janzen2001}. The electron spin of the isolated paramagnetic defects in SiC can serve as a solid-state qubit once it can be initialized, read out and coherently controlled~\cite{Gali2010, Christle2015, Widmann2015, Christle2017, WangPRL2020, Wolfowicz2020, Cilibrizzi2023}. Furthermore, nuclear spins of the host crystal proximate to the point defects, e.g., $^{13}$C and $^{29}$Si in SiC, or introduced by the defect itself, e.g., $^{14}$N or $^{15}$N in nitrogen-vacancy defects~\cite{Gordon2015, Bardeleben2016, Csore2017, WangPRL2020, WangACSPhotonics2020, Mu2020, Murzakhanov2021, Bardeleben2021, Narahara2021} or $^{51}$V in substitutional vanadium defects~\cite{Schneider1990, Kaufmann1995, Baur1997}, can be used as quantum registers or ancilla qubits mediated by the hyperfine interaction between them (e.g., Refs.~\onlinecite{Falk2015, Ivady2016, Bourassa2020}). The quantum states of the electron spin and nuclear spins can be typically controlled by microwave and radio-frequency alternating magnetic fields, if necessary, in the presence of a constant magnetic field. 

A coherent spin-photon interface (SPI) can be created by exploiting spin-dependent atomic-like optical transitions~\cite{High-fidelity_Read-out_2011, Christle2017, Nagy2019, Cilibrizzi2023}, typically available at cryogenic temperature. Spin-selective optical transitions, possibly in combination with microwaves for higher-spin systems, enable the preparation of specific spin states by optical pumping~\cite{Nagy2019, Tissot2022}. They also enable readout by detection of the emitted photons under optical excitation~\cite{High-fidelity_Read-out_2011}, or by conversion of spin-states into different charge states~\cite{rogge_Er_SSRO2013, zhang_SSRO2021, anderson_5s_SSRO_SiC2022}. Coherent SPI's also enable to establish a connection between the quantum memory of the solid-state qubit and the flying qubit encoded into the chosen quantum property of the photon~\cite{Christle2017}. This can be harnessed to setup entanglement-based quantum communication.
Recently, the first demonstration of spin-photon entanglement with a SiC defect was reported using the silicon vacancy in 4H-SiC~\cite{Fang2023}. 

Architectures based on single point defects in materials with a diluted nuclear spin bath have also shown the promising opportunity to coherently control a register of nearby weakly-coupled nuclear spins~\cite{Taminiau12}. In this case, the hyperfine interaction acts on timescales longer than the electron spin dephasing time, so that it does not result in a separate transition visible in the electron spin resonance spectrum and cannot be driven directly. Nonetheless, it is possible to address and coherently control such nuclear spins by applying pulse sequences such as dynamical decoupling on the electron spin, that extend its coherence time and can effectively isolate the contribution of a single, weakly-coupled nuclear spin, filtering out all the other background. These techniques have enabled the implementation of multi-qubit schemes~\cite{taminiau_PRX_10spins}, such as quantum error correction~\cite{cramer_repeated_2016}. While most of this work has been done in diamond, some recent results show that it is also possible in SiC~\cite{Bourassa2020}. Given the higher abundance of isotopes with non-zero nuclear spin in SiC, however, isotopic engineering may be required to maximize the number of controllable nuclear spins~\cite{nagy_PrAppl2023}.

We note that the reduction of the density of $^{13}$C and $^{29}$Si nuclear spins can be advantageous to extend the coherence times and reduce the optical linewidth of the coherent emission~\cite{Bourassa2020, Cilibrizzi2023}, which occurs at the zero-phonon-line (ZPL) emission of the defects. It is imperative to fully characterize the magneto-optical properties of these point defects, in order to develop efficient quantum optics protocols for quantum information processing applications such as quantum communication.

\begin{figure*}
    \centering
    \includegraphics[width=1\textwidth]{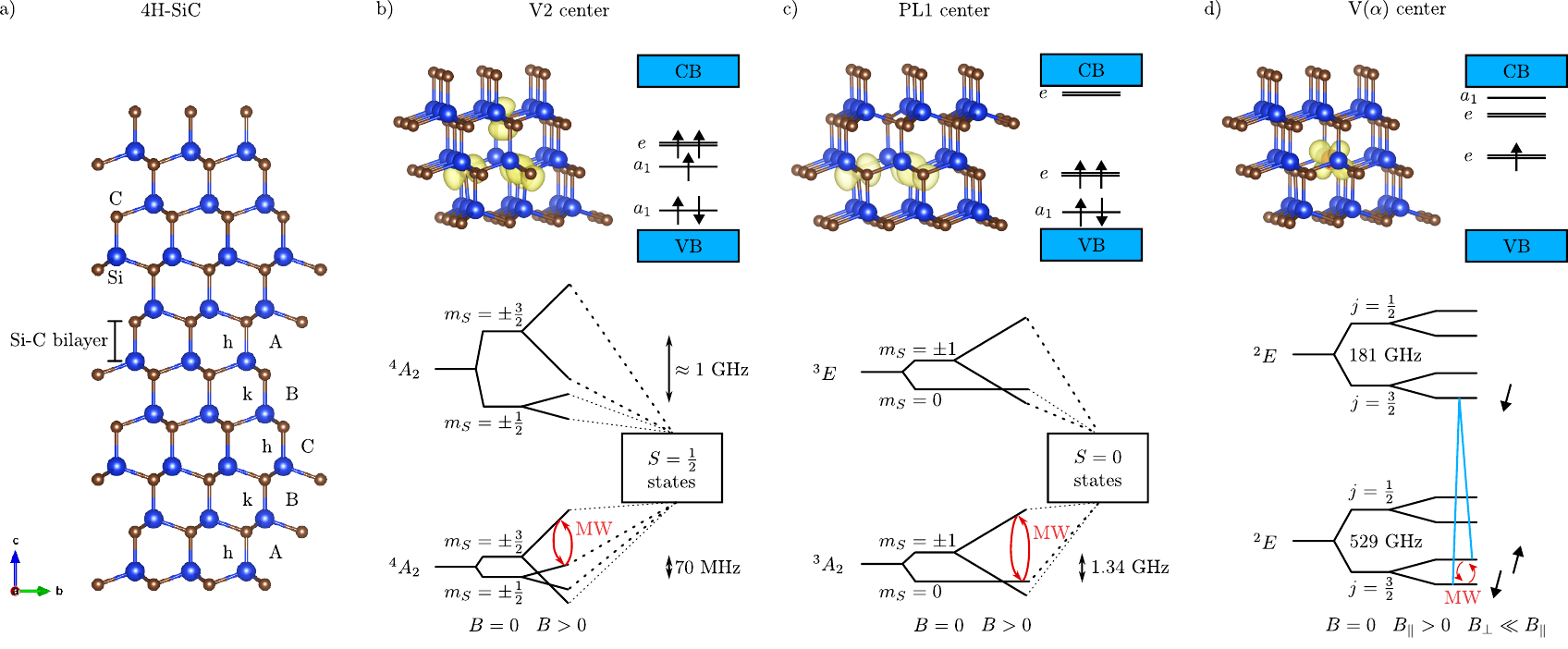}
    \caption{\label{fig:defect_molecule}(a) 4H-SiC crystal lattice with its stacking sequence, (b) Silicon-vacancy and its electronic structure in the negative charge state at the quasicubic site, (c) the neutral divacancy with a neighboring Si and C vacancy with its electronic structure and (d), Vanadium substituting Si and its electronic structure in the neutral charge state at the quasicubic site.}
\end{figure*}

\CB{The most mature polytype of SiC is 4H-SiC which has a band gap of 3.3~eV. To realize a spin-to-photon interface, deep levels should be introduced by the point defect. This can typically be achieved by introducing dangling bonds of vacancies or dopants with $d$ and $f$ orbitals. We illustrate this by comparing three types of deep-level point defects in SiC: the silicon-vacancy (Si-vacancy), divacancy and vanadium centres in 4H-SiC (see Fig.~\ref{fig:defect_molecule}). These defects introduce multiple levels into the band gap where the excited state can be described by promoting an electron from the occupied in-gap level to the unfilled in-gap level. These defects have complementary ground state spin structure, with the the negatively charged Si-vacancy having spin $S=3/2$ in an orbital singlet, the neutral divacancy spin $S=1$, and the neutral vanadium substituting silicon has spin $S=1/2$ in an orbital doublet.} 

The electron-orbital and spin level structures 
are depicted for the three selected qubit species (see Fig.~\ref{fig:defect_molecule}).
For the negatively charged Si-vacancy in 4H-SiC, in-gap localized defect states are generated by the four carbon dangling bonds of the defect which lead to non-degenerate and triple-degenerate orbitals occupied by five electrons. The latter splits due to the $C_{3v}$ symmetric crystal field of the defective 4H-SiC lattice. Finally, these defects states constitute an orbital singlet $^4A_2(g)$ ground state and an orbital triplet $^4T_2(e)$ excited state, where the latter splits to $^4A_2(e)$ and $^4E(e)$ states in ascending order for the reason given above~\cite{Janzen2009}. We note that the strength of the crystal field and the energy spacing between these levels depends on the actual defect site and in 4H-SiC the splitting is larger for the Si-vacancy defect at the quasicubic site~\cite{Ivady2017}. We continue with that Si-vacancy configuration in 4H-SiC. At low temperatures, the fluorescence is dominated by the ${}^4A_2(e)\rightarrow{}^4A_2(g)$ optical transition which results in the so-called V2 PL spectrum with zero-phonon-line (ZPL) wavelength at 917~nm (1.35~eV)~(see Ref.~\onlinecite{Janzen2009} and references therein). The $m_S=\pm1/2$ and $m_S=\pm3/2$ states of the quartet spin slightly split by the so-called zero-field-splitting of about 70\,MHz due to the crystal field. With optical excitation of this system, the electron spin can be pumped into the $m_S=\pm1/2$ in the electronic ground state, and the photoluminescence is brighter for the $m_S=\pm1/2$ states than that for $m_S=\pm3/2$ states. The origin of the spin-selective fluorescence comes from the spin-selective intersystem crossing between the $^4A_2(e)$ state and the series of spin doublet excited states lying between the quartet states~\cite{Baranov2011, Gali2012, Riedel2012, Soltamov2012, Soykal2016, Soykal2017, Dong2019, Widmann2019, Wang2021}, see also Fig.~\ref{fig:defect_molecule}(b). At zero external magnetic field, radiofrequency magnetic fields can rotate the electron spin in the ground state manifold. This electron spin resonance frequency can be increased to the microwave regime upon applying an appropriate external magnetic field aligned with the c-axis (defect symmetry axis or crystal stacking axis, see Fig.~\ref{fig:defect_molecule}(a)) of 4H-SiC.

The neutral divacancy~\cite{Son2006, Gali2010, Gali2011, Koehl2011, Ivady2019, Li2022} in SiC consists of neighbouring Si and C vacancies in the SiC lattice [see Fig.~\ref{fig:defect_molecule}(c)]. The electronic structure is identical to the NV centre in diamond, with C$_{3v}$ symmetry and a $S=1$ electronic spin, so that spin-conserving cyclic transition between $^{3}A_2 \rightarrow ^{3}E$ can be used for spin-photon interfacing~\cite{Christle2017}. Difference from the NV centre are the emission wavelength in the near-infrared (with zero-phonon lines around 1030-1130nm) and a zero-field splitting of about $1.3$ GHz. Experimental progress on the divacancy has been quite fast with several breakthroughs, such as long spin coherence time, single-shot electron spin readout~\cite{anderson_5s_SSRO_SiC2022} and control of associated nuclear spins~\cite{Bourassa2020} demonstrated in just a few years. 

Another configuration of interest is an electron with $S=1/2$ within a doublet orbital state. This provides altogether \(2 \times 2\) quantum states that split into two doubly degenerate Kramers states due to spin-orbit coupling.
The Kramers degeneracy can be further split by magnetic fields. 
Neutral vanadium substituting the silicon in the SiC forms such a system.
The in-gap defect states come from the vanadium $d$ orbitals, which are five-fold degenerate under spherical symmetry. The crystal field provided by the SiC lattice splits these $d$ orbitals into a low-energy orbital doublet -- which interacts weakly with the lattice -- and an orbital triplet higher in energy that hybridizes with the neighboring carbon orbitals. This orbital triplet is further split into a doublet and a singlet due to the $C_{3v}$ symmetry of the crystal field.
In the neutral vanadium center the lower-energy double degenerate in-gap state is occupied by a single electron which manifests the $^2E(g)$ ground state. The strength of the spin-orbit splitting in the $^2E(g)$ state depends on the specific defect site which is as large as 529~GHz for the quasicubic configuration. The orbital triplet $^2T_1(e)$ excited state splits to $^2E(e)$ and $^2A_1(e)$ levels where the order of these levels also depend on the defect site; the ascending order is $^2E(e)$ and $^2A_1(e)$ at the quasicubic site~\cite{Spindlberger2019}. The optical transition can be described as ${}^2E(e)\rightarrow{}^2E(g)$ which is associated with the $\alpha$ vanadium photoluminescence (PL) center in 4H-SiC~\cite{Baur1997, Spindlberger2019}. As the excited state is less localized than the ground state and the excited state partially loses its $d$ orbital character, the excited-state lifetime is on the order of 100~ns~\cite{Spindlberger2019}.  

A similar electronic structure and optical transition ${}^2E_u(e)\rightarrow{}^2E_g(g)$ is known for the negatively charged silicon-vacancy PL center~\cite{Sternschulte1994, Goss1996, Neu2011, Gali2013, Hepp2014} and akin to group-IV--vacancy PL centers~ (see Refs.~\onlinecite{Thiering2018, Ruf2021} and references therein) in diamond: Resonant excitation between the selected states in the ground state and optical excited state manifolds results in coherent population trapping, and a so-called $\Lambda$-scheme can be used to control the qubit state~\cite{Pingault2014, Rogers2014}. The most developed realizations of the SPI have been demonstrated for the negatively charged silicon-vacancy PL centers among the diamond group-IV--vacancy PL centers~\cite{Becker2016, Pingault2017, Sukachev2017, Becker2018, Sohn2018, Machielse2019, Bhaskar2020}. This center has a ZPL at 738~nm (1.68~eV) which requires a frequency conversion to telecommunication wavelengths for efficient entanglement distribution over optical fibers~\cite{Bersin2023}. 
Based on the similarities in the nature of optical transitions of the diamond silicon-vacancy PL center and the $\alpha$ vanadium PL center, several techniques can be transferred to realize the SPI [see Fig.~\ref{fig:defect_molecule}(d)]. 
It comes with a significant advantage over the diamond group-IV--vacancy centers, as the $\alpha$ center's ZPL at 1278~nm (0.97~eV) falls into the telecommunication O-band, removing the need for complex and inefficient nonlinear wavelength conversion systems. We note that standard SMF-28 telecom fibers enable single-mode propagation both for O- and C-band photons: this is practically very convenient as the (largely unoccupied) O-band can be used for the single-photon-level quantum channel, and the C-band for the required classical channel, avoiding cross-talk. 

Single-photon emission from vanadium PL centers in SiC has been observed~\cite{Wolfowicz2020, Cilibrizzi2023}, where the interpretation of the fine structure PL requires 
the details of the vanadium hyperfine interaction with the electron spin and the double degenerate orbital~\cite{Tissot2021, Gilardoni2021}. 
The interaction of the magnetic field with this system is a complex interplay between the electron-phonon coupling, spin-orbit coupling and the specific selection rules of the $d$ orbitals of vanadium~\cite{Csore2020}. 
Understanding the fine details of the spin Hamiltonian is a necessary step to design quantum optics protocols for optical control of the vanadium nuclear spin~\cite{Tissot2021, Tissot2022, Astner2022}. 
The observed $T_1$ time reaches 25~s at 100~mK~\cite{Astner2022} which enabled to characterize the hyperfine level structure of the $\alpha$ center via two-photon magneto-spectroscopy and optically detected magnetic resonance~\cite{koller2025strainenabledcontrolvanadiumqudit}.
For ensemble $\alpha$ centers, the observed spin dephasing time $T_2^*$ from Ramsey interference measurements is 7.2~$\mu$s at 2~K (Ref.~\onlinecite{Hendriks2022}). The observed 150~counts/s from single vanadium emitters~\cite{Cilibrizzi2023} may be substantially improved by engineering the single vanadium defects into solid immersion lenses~\cite{bekker_SILs_2023}, photonic crystals
or optical waveguides and open microcavities~\cite{Bracher2015, Radulaski2017, Babin2022}. These parameters and properties make the $\alpha$ vanadium center in 4H-SiC very promising candidate to realize a practical SPI with a programmable nuclear spin quantum memory.      

We note that the single vanadium centers have been created in SiC by vanadium ion implantation, followed by high-temperature annealing to repair lattice damage of the crystal lattice~\cite{Wolfowicz2020, Cilibrizzi2023}. The residual parasitic defects after annealing may create strain and fluctuating charge environment upon illumination. This is an important issue in securing stable emission from single vanadium centers and producing indistinguishable quantum emitters for quantum communication. Vanadium in isotopically purified 4H-SiC has a remarkably narrow distribution of central emission frequencies. However, the linewidth of the emission remains several orders of magnitude larger than the Fourier limit~\cite{Cilibrizzi2023}. This broadening is believed to be caused by charge flctuations in the vicinity of the defect under illumination. The fluctuating charges may be depleted by integrating the vanadium centers into p-i-n junctions of 4H-SiC and applying an appropriate electric field, a method which is already established for isolated divacancy qubits~\cite{Anderson2019}. A recent study on single vanadium centers in 4H-SiC has shown that the photostability of single $\alpha$ centers upon illumination can be significantly improved by compensating the residual n-type and p-type dopants of SiC and pinning the Fermi-level to the middle of the band gap~\cite{Cilibrizzi2023}. 
Further spin centres in SiC have recently been studied which possess bright ZPL emission in the telecom domain, including vanadium in 3C-SiC at 1495 nm and the family of chlorine-related defects, with ZPL wavelengths between 1330 nm and 1590 nm~\cite{Shafizadeh2026vanadium3C,Anisimov2025chlorine}, and much remains to be discovered regarding the properties of these and other, as yet unknown, spin centres in SiC.

\subsection{SiC photonics} \label{sec:sicdevices}

SiC is a mature platform for microelectronics, in particular for power applications. Capitalizing on a wider bandgap, higher thermal
conductivity, and larger critical electric field than silicon, SiC devices can operate at higher temperatures, higher current density, and
higher blocking voltage and are becoming more and more widespread in applications such as power conversion in electric vehicles~\cite{she2017review,kimoto2022high}

Recently, researchers have begun to explore SiC applications beyond power devices~\cite{LaVia2023, Ou2023}. While it is not a material widely utilised in photonics applications yet, it exhibits rather promising optical properties~\cite{Castelletto2022}. Due to its wide bandgap, SiC presents a wide transparency window, between 0.37 - 5.6 $\mu$m for the 4H polytype~\cite{Wang2013midIR, zheng20194h, zheng2019high}.

SiC also features fairly strong nonlinear coefficients, not far from values reported for silicon and lithium niobate. The quadratic nonlinearity $\chi^{(2)}$ has been measured to be $\approx$ 12.5 pm/V in the hexagonal polytypes~\cite{sato2009accurate}, while the third order (Kerr) nonlinearity $\chi^{(3)}$  is about 6.9$\times 10^{- 19}$m$^2$/W at 1550 nm~\cite{Guidry_parametric_optica_2020} for 4H-SiC. Strong nonlinear coefficients are crucial to implement on-chip modulators and frequency converters~\cite{Shi2023,  Shi2021, Yi2020, Zheng2019}.

Given its outstanding electronic and photonic properties, SiC provides a formidable platform for integration of different functionalities into the same chip. It is possible to envision a single quantum photonic chip integrating SPI's with all photonic elements required to process quantum states of light (such as optical modulators or beamsplitters), all electrically controlled.

One crucial step for high-rate spin-photon interfacing is to enhance light-matter interaction with an optical cavity. Enhancement of light-matter interactions using cavity quantum electrodynamics (CQED) can broadly be divided into two approaches: on the one hand, cavities can increase the photon emission from defects into a desired spatial mode, on the other hand a cavity can enhance interaction of an incoming light pulse with the optical dipole. We will briefly describe both approaches below.

The enhancement factor for the spontaneous emission into the cavity, $F_{SE}$ is given by

\begin{equation}
    F_{SE}=\frac{3}{4\pi^2}\Upsilon\frac{\gamma_{ZPL}}{\gamma_{tot}}\zeta^2,\text{with } \Upsilon=\left(\frac{\lambda}{n}\right)^3\frac{Q}{V},
\end{equation}\label{eq:cavityEnhancement}

with the free-space wavelength $\lambda$, the refractive index of the material $n$, the quality factor $Q$, and the mode volume $V$. We have furthermore defined the optical enhancement factor $\Upsilon$ which collects the performance parameters of the cavity. The value $\zeta\in[0,1]$  gives the overlap of the defect's dipole with the electric field mode, relating both to the relative orientation of the dipole and its position within the spatial distribution of the mode, while $\gamma_{ZPL}$ and $\gamma_{tot}$ are the decay rate into the zero-phonon line and the total decay rate, respectively~\cite{Chatzopoulos2019, Faraon2013}.

Further, an incoming light pulse will interact more strongly with the optical dipole transition of the defect if it is placed in a suitable cavity. If transitions of different spin states can be resolved, the latter mechanism can be implemented for spin-dependent absorption, reflection, or phase shifting of the incoming light. For a symmetric, lossless cavity, the fractions of the resonant optical power reflected and transmitted by the coupled system are given by~\cite{Reiserer2015}
\begin{equation}
    R_C=\left(\frac{2C}{2C+1}\right)^2\,\text{ and }\,
    T_C=\left(\frac{1}{2C+1}\right)^2,
\end{equation}\label{eq:cavityReflTrans}

where $C=F_{SE}/2$ is the cooperativity of the system. For finite cooperativity, a fraction of the light will be scattered and lost by the system and is given by $S_C=4C/(2C+1)^2$. Such scattering processes can further lead to undesired spin flips. These expressions assume low-power excitation, i.e. far below the saturation power of the system.

The last few years have seen strong progress in the development of SiC microcavities~\cite{Lukin2020, Castelletto2022}. An important starting point is the fabrication of thin SiC membranes. Few-microns thick membranes on a high-reflectivity Distributed Bragg Reflector (DBR) can be used for open microcavities, positioning a second concave DBR on top of the emitter~\cite{Fait2021}. Thinner suspended membranes (few hundred nanometers thick) are required to fabricate photonic crystal devices such as photonic crystal nanocavities and nanobeams. Seminal papers have demonstrated the fabrication of membranes and cavities~\cite{Bracher2015, bracher_selective_2017, Crook2020,mokhtarzadeh2022} by electro-chemical etching, exploiting the etching selectivity between regions of different doping. More recently, researchers have developed thin-film SiC on insulator, by bonding SiC on a SiO$_2$-on-Si wafer, and grinding/polishing the SiC to the desired thickness~\cite{Lukin2020}. This technique can deliver wafer-scale thin-film SiC with low roughness, and has been used to fabricate different types of microcavities~\cite{yang_inverse-designed_2023, lukin_two-emitter_2023}. However, such processes are currently hampered by non-uniformity of the SiC thickness, leading to low yield in photonic device fabrication. 

SiC photonic crystal cavities have reached extremely high quality factors of $6.3\times10^5$ while maintaining a small mode volume of $2.1\left(\lambda/n\right)^3$, corresponding to $\Upsilon=3\times 10^5$~\cite{Song2019}.
While integration of SiC with Fabry-Pérot resonators at telecom wavelengths remains unreported, recent work has demonstrated SiC membrane-based microcavities at a different wavelength regime~\cite{hessenauer2025cavityenhancementv2centers}, indicating promising potential for future development.
In the telecom domain, a Fabry-Pérot microcavity containing an yttrium orthosilicate (YSO) membrane reached $\Upsilon=7\times10^3$~\cite{Merkel2020}. Recently, microcavities with smaller mode volume and higher finesse have been demonstrated, indicating that  $\Upsilon=2\times10^4$ could be reached for SiC~\cite{Fait2021}.
While the enhancement factor is expected to be lower than for optimized photonic crystal cavities, Fabry-Pérot microcavities offer highly efficient coupling to single-mode fibers and integration with micro-electro-mechanical systems for individual tuning in large-scale systems~\cite{Derntl2014}.

From a combination of measurements and \emph{ab-initio} calculations, vanadium on the $\alpha$ site of 4H SiC is expected to have a branching ratio of $\gamma_{ZPL}/\gamma_{tot}\simeq9\,\%$~\cite{Spindlberger2019}, giving $C$ approaching 100 for the best reported Fabry-Pérot microcavities, and even approaching 1000 for the best reported photonic crystal cavities, respectively~\cite{Fait2021,Song2019}. In both cases, high contrast between coupled and uncoupled spin states can be expected. 

\section{Beyond point-to-point quantum key distribution with SiC devices}

In the following, we focus on applying SiC devices in quantum key distribution (QKD), which is arguably the most mature family of quantum communication protocols.
This approach is particularly useful, since the performance of SiC devices in quantum networks can now be assessed and optimized based on a single parameter, namely the secure key rate (SKR) in QKD, which is the rate of distributed bits that can be used for encryption purposes.
The SKR can be used as a key performance metric for quantum networks, as it depends not only on the achievable photon throughput but also on errors arising during transmission and imperfections inherent to the devices. 
In direct transmission, the SKR is limited by the photon repetition rate of the source and the time resolution of the detection system.
In memory-enhanced quantum networks, processing time, communication latency, and quantum node efficiency additionally limit the SKR, along with system noise, as will be discussed in this chapter.

In QKD, two parties, Alice and Bob, aim to establish a secret key with information-theoretic security. 
Alice encodes raw key in qubits prepared in randomly chosen conjugate bases, while Bob measures them using bases randomly selected from the same set. 
They retain the secure key bit from events where their basis choices match. 
Due to the no-cloning theorem~\cite{park1970concept, wootters1982single}, a third party is fundamentally incapable to obtain the key without prior knowledge of Alice's or Bob's basis settings. 
Any eavesdropping attempt disturbs the quantum state, revealing an adversary.
There are two families of QKD protocols. 
In the prepare-and-measure type (e.g. the BB84 protocol~\cite{bennett2014quantum}), Alice prepares the quantum states and sends them to Bob, who performs measurements. 
Entanglement-based protocols (e.g. the BBM92 protocol~\cite{bennett1992quantum}) consist of a central node which distributes pairs of entangled photons to Alice and Bob, both of which perform measurements on their received photon. 

Practical implementations of QKD over long distances are mainly impeded by photon loss in optical fibers.
The loss of photons leads to an exponential decrease of the SKR with fiber distance, limiting the maximal key exchange distance to a few hundred kilometers in optical fiber~\cite{neumann2022continuous}. 
This is the case for a direct fiber connection between Alice and Bob. 
An alternative approach is the division of the total distance between Alice and Bob into several segments. 
The loss in each of these segments is reduced exponentially, leading to a substantial improvement in the SKR between Alice and Bob.
There are two fundamentally different approaches which make use of this segmentation. 
The first approach is known as a trusted repeater, where each of the nodes connecting two segments performs a full QKD protocol and the resulting key is relayed to other segments.
While this indeed increases the possible transmission distance indefinitely, each node reads out the key, which necessitates absolute trust in each node.
The second approach is known as a quantum repeater~\cite{Briegel1998, Azuma2023}.
Here, adjacent nodes are supplied with entangled photons which are stored in quantum memories. 
Once the quantum memories of all nodes are loaded, nested entanglement swapping is performed in order to entangle the end nodes which belong to Alice and Bob.  
With entanglement at hand, Alice and Bob can now perform any quantum communication protocol using entanglement as a resource~\cite{Horodecki2009a}.

SiC devices are natural candidates for this purpose, since quantum repeater stations must facilitate the storage of qubits over long periods of time and process stored qubits for entanglement distillation~\cite{bennett1996purification}.   
Despite their potential, current SiC devices, as well as other quantum memory platforms, are unable to fulfill the stringent requirements of quantum repeaters, let alone the necessary error correction or entanglement distillation overheads, and therefore fail to outperform direct transmission of photons. 
However, there are QKD protocols which exhibit a quantum memory enhancement without requiring a fully-fledged quantum repeater chain~\cite{Azuma2022,Rozpdek2019}. 

We focus on one particular protocol, so-called memory-assisted measurement-device-independent QKD (MA-MDI-QKD)~\cite{Panayi2014}, which requires only a single quantum repeater node. 
First, we cover the deployment of SiC devices as quantum nodes and discuss to what extent they fulfill the requirements for quantum communication.
We then go through the MA-MDI-QKD protocol in greater detail and simulate the protocol based on realistic parameters of state-of-the-art SiC devices. 
Additionally, we discuss the opportunities of SiC devices for advanced quantum network architectures.

\subsection{SiC devices as quantum nodes} \label{sec:qunode}
\begin{figure}
    \centering
    \includegraphics[scale=0.6]{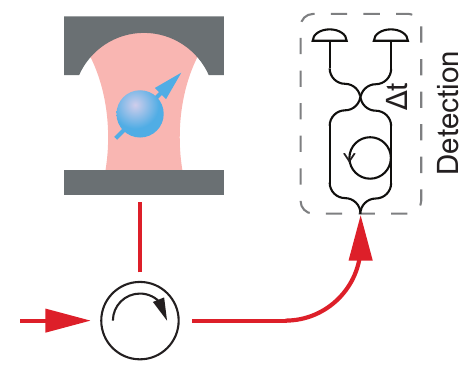}
    \caption{
    Configuration of a CQED device enabling a heralded quantum memory and an asynchronous Bell state measurement (BSM). 
    The time-bin state of an incoming photon is converted into spin-photon entanglement between the electron spin and an outgoing time-bin encoded photon. 
    A measurement of the reflected photon in an imbalanced Mach-Zehnder interferometer with imbalance $\Delta t$ transfers the state of the initial photon onto the spin state, heralded by a single-photon detection event.
    }
    \label{fig:memorywriting}
\end{figure}

SiC devices serve as an SPI, enabling the transfer of a photon's quantum state to an electronic spin state and vice versa.
A device configuration implementing this interface is sketched in Fig.~\ref{fig:memorywriting} and discussed below.
After the qubit is stored in the spin state, we can apply arbitrary unitary operations using microwave pulses. 
To connect two quantum links, we need to perform at least two-qubit gates in the quantum node.

The SPI primarily relies on spin-photon entanglement. 
Let us consider the writing process of a photonic qubit state into the quantum memory. 
We use time-bin encoding of the photon, although it is worth noting that SiC memories can also be connected to other degrees of freedom, such as polarization.
A time-bin qubit 
\begin{equation}
\label{eq:time-bin-qubit}
    \ket{\Psi}_\text{photon} = \frac{1}{\sqrt{2}} \left(\ket{t_1} + \text{e}^{i\phi}\ket{t_2}\right),
\end{equation}
is realized as a coherent superposition of two orthogonal states, corresponding to the presence of a photon at time $t_1$ and at a later time $t_2$.  
The electron spin is initialized in state 
\begin{equation}
    \label{eq:spin-init}
    \ket{\Psi}_\text{spin}^\text{init}=\frac{1}{\sqrt{2}}\left(\ket{\downarrow}+\ket{\uparrow}\right),
\end{equation}
where $\ket{\downarrow}$ ($\ket{\uparrow}$) corresponds to the spin down (up) state of the electron. 
Based on the non-degenerate level structure~\cite{nguyen2019quantum}, a photon arriving at time $t_1$ is resonant only with the optical transition of state $\ket{\downarrow}$ and is thus reflected off the defect-cavity system via coherent scattering, resulting in the joint spin-photon state $\ket{\downarrow t_1}$.
Between $t_1$ and $t_2$, a $\pi$-pulse is applied to the microwave transition, which inverts the electron spin state swapping $\ket{\downarrow}$ and $\ket{\uparrow}$. 
The later time-bin $\ket{t_2}$ is again resonant with the optical transition of state $\ket{\downarrow}$, and upon successful reflection and another $\pi$-pulse results in state $\ket{\uparrow t_2}$.
Both the population transfer between the electronic spin states and the time-bin state are coherent processes, resulting in a coherent superposition of spin-photon states.
\begin{equation}
    \ket{\Psi}_\text{spin-photon} = \frac{1}{\sqrt{2}}\left(\ket{\downarrow t_1}+\text{e}^{i\phi}\ket{\uparrow t_2}\right),
\end{equation}
describing a spin-photon-entangled state.

An optical circulator directs the reflected photon to a detection unit, comprising a Mach-Zehnder interferometer with imbalance $\Delta t = t_2 - t_1 $ and single-photon detectors (see Fig. \ref{fig:memorywriting}). 
The interferometer implements a probabilistic measurement in the X-basis $\{\left(\ket{t_1}+\ket{t_2}\right)/\sqrt{2},\left(\ket{t_1}-\ket{t_2}\right)/\sqrt{2}\}$.  
Detection of the photon in this basis transfers the initial photonic qubit state onto the spin state, resulting in
\begin{equation}
    \ket{\Psi}_\text{spin}^{\text{final}} = \frac{1}{\sqrt{2}}\left(\ket{\downarrow}+\text{e}^{i\phi}\ket{\uparrow}\right),
\end{equation}
where the measurement-induced phase ambiguity, resulting from the probabilistic registration in one of the two single-photon detectors after the Mach-Zehnder interferomter, has already been compensated \footnote{Without loss of generality we assume that the compensation is performed immediately after the photon measurement. 
In a quantum communication setting, the compensation will not be performed immediately since unitary operations on the spin are time-consuming. Instead, the measurement outcome is stored and classically communicated at a later time.}. 

The detection of a single photon after the interferometer heralds the successful writing process of the initial photon state to the spin state memory.
The overall efficiency of the described writing process is limited to a maximum of \SI{25}{\percent}, stemming from two probabilistic steps: the state-dependent photon reflection at the cavity succeeds only in \SI{50}{\percent} of the cases, and the projective measurement in the Mach-Zehnder interferometer also succeeds only in \SI{50}{\percent} of the cases~\cite{Franson1989}.
This photon loss after interaction with the spin-cavity system not only reduces the SKR due decreased signal intensity but also introduces noise at the memory loading stage~\cite{Sohr2026_unpublished}.

Two-qubit gates play a crucial role in quantum communication, especially for achieving a BSM between two incoming photons. 
SiC devices provide a suitable approach to perform this BSM asynchronously using a single defect~\cite{Bhaskar2020}. 
After writing the first photon in the spin memory, a second photon with phase $\phi'$ can be loaded into the same defect memory. 
The resulting spin state 
\begin{equation}
    \ket{\Psi}_\text{spin}^{\text{BSM}} = \frac{1}{\sqrt{2}}\left(\ket{\downarrow}+\text{e}^{i(\phi+\phi')}\ket{\uparrow}\right),
\end{equation}
inherits the sum of the phases of both photons. 
Once more, the phase ambiguity that arises after photon measurement has already been compensated for.
The BSM is concluded by reading out the spin state in the X-basis. 
Depending on the measurement outcome, two Bell states, namely the $\Phi^+$ or the $\Phi^-$ state, can be discriminated.
While the asynchronous BSM is resource efficient -- only a single quantum memory is required -- both photons enter the cavity via the same single-mode fiber. 
To prevent additional channel losses caused by probabilistic routing with a beamsplitter~\cite{Bhaskar2020}, active optical switching can be implemented instead.

\subsection{Memory-assisted measurement-device-independent QKD} \label{sec:MAQKD}
\begin{figure}
    \centering
    \includegraphics[width=1\linewidth]{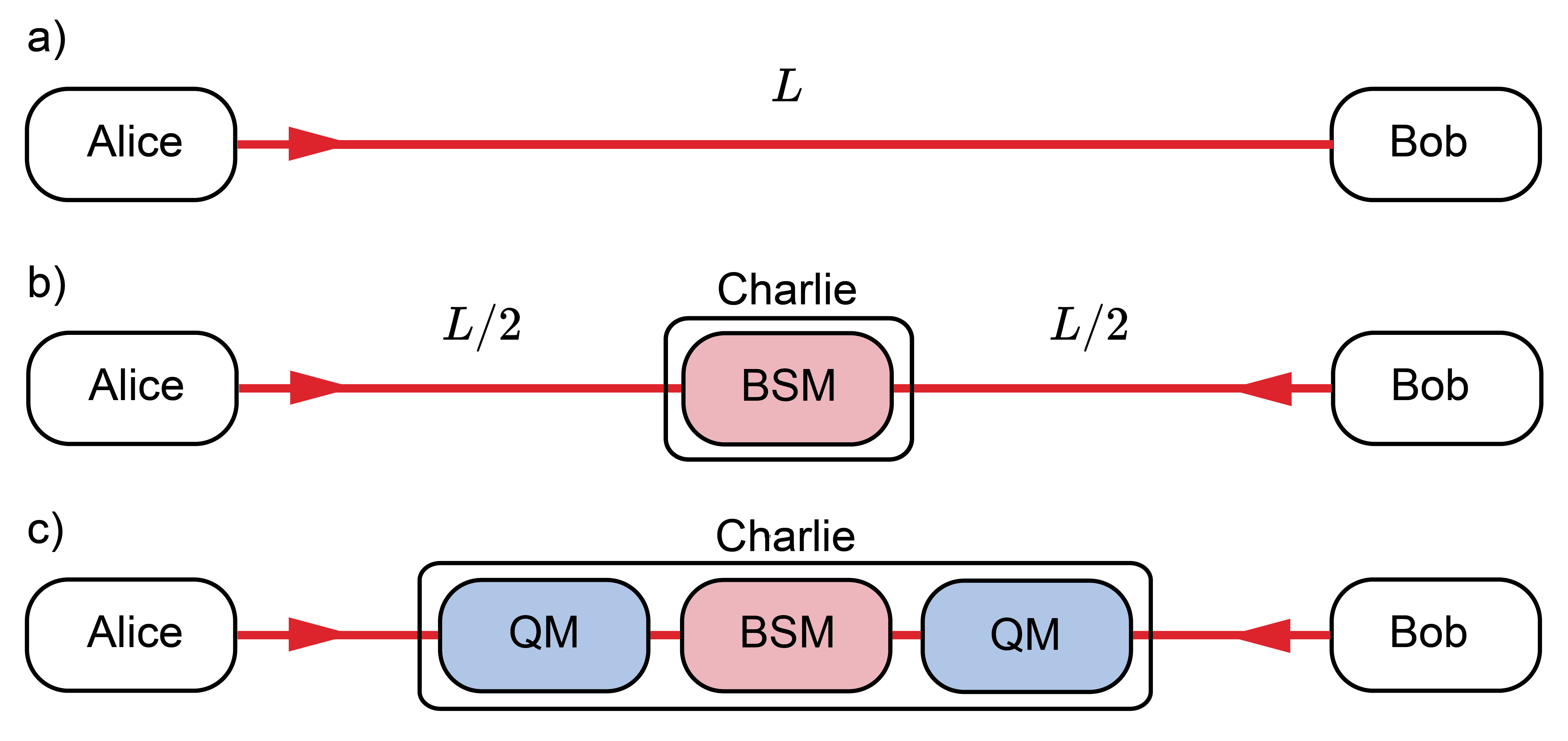}
    \caption{Comparison between different QKD schemes. a) In prepare-and-measure-type QKD protocols (e.g., BB84), Alice produces quantum states and sends them to Bob over a channel of length $L$. Bob performs measurements. b) In measurement-device-independent (MDI)-QKD protocols, Alice and Bob each prepare quantum states and send them to Charlie, an untrusted third party located at the channel midpoint (channel segment length $L/2$). Charlie performs a Bell state measurement (BSM) on both quantum states and based on the outcome of this measurement, Alice and Bob agree on a secret key. c)  In memory-assisted measurement-device-independent (MA-MDI)-QKD, both Alice and Bob produce quantum states and send them to Charlie. Prior to the BSM, Charlie synchronises the arrival of the quantum states with the help of quantum memories (QM). Upon successful BSM, Charlie announces the outcome of the measurement which leads to the establishment of a secret key between Alice and Bob.}
    \label{fig:MA_QKD_scheme}
\end{figure}

Measurement-device-independent (MDI) QKD can be viewed as an entanglement-based QKD protocol in reverse. 
In entanglement-based QKD a central source produces pairs of entangled photons and transmits them to the communicating parties Alice and Bob. 
In MDI QKD, both Alice and Bob randomly and independently prepare photons in one of the four BB84 states \{\(\ket{0}\), \(\ket{1}\), \((\ket{0}\pm\ket{1})/\sqrt{2}\)\}~\cite{bennett2014quantum} and send them to a third, untrusted party, called Charlie (C) (see Fig.~\ref{fig:MA_QKD_scheme}).
Charlie performs a BSM, projecting the separable state of Alice's and Bob's photon on a joint state in the Bell basis. 
Following Charlie's announcement of the BSM outcome, Alice and Bob publicly disclose their basis choices, retaining only instances where their bases match and discarding the rest.
Based on Charlie's announcement, Alice or Bob adjust their bit values as necessary to ensure correlation of their bit strings.

This protocol is measurement-device-independent, since it avoids known side-channel attacks and vulnerabilities of the detectors~\cite{Lo2012}.
The protocol requires that both Alice's and Bob's photon arrive simultaneously at the BSM. 
With increasing channel loss, the probability of two photons arriving simultaneously decreases exponentially.

The efficiency of this protocol can be significantly bolstered by temporarily storing qubits at the central node prior to the BSM. 
Such storage enables synchronisation of Alice's and Bob's channels~\cite{Panayi2014}, eliminating the need for precise simultaneous photon arrival. 
The BSM is then executed asynchronously, exclusively upon the complete loading of both memories.
In essence, this scheme merges ideas from quantum repeaters and MDI-QKD and is among the simplest QKD protocols involving quantum memories. We discuss this protocol as it provides a straightforward performance comparison to direct photonic QKD systems. More advanced methods may allow to further increase the quantum communication rates using scalable spin-photon interfaces~\cite{Muralidharan2016,Wo2023,Pettersson2025}.

\begin{figure}
    \centering
    \includegraphics[scale=0.6]{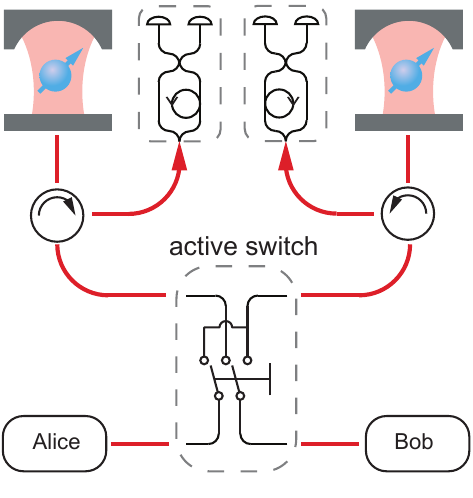}
    \caption{Two CQED devices receiving photons from Alice and Bob via an optical switch. Each of the devices can perform asynchronous Bell state measurements. In order to avoid idle time of the central node, the optical switch toggles Alice's and Bob's fibers after successful detection of one photon either from Alice or Bob.}
    \label{fig:QKD_schematic}
\end{figure}

The asynchronous BSM introduced in the previous section is perfectly suited for MA-MDI-QKD  with minor modifications.
Alice and Bob each randomly and independently choose their basis as well as their bit value.
They prepare time-bin states, such as described in Equation~(\ref{eq:time-bin-qubit}) and encode their bit choice in the relative phases $\left\lbrace0,\pi\right\rbrace$ and $\left\lbrace \pi/2,-\pi/2\right\rbrace$ for the $X$ and the $Y$ basis respectively.
The qubits follow one after another with the same temporal spacing as the two time-bins forming a qubit. 
Neither do qubits interleave, that is, share the same time-bin, nor is there extra spacing between successive qubits.
Since both Alice's and Bob's photons must be guided to the same SPI and the sender of each photon must be unambiguously identified, optical switching is required.
An efficient solution to this problem is to use two SiC devices and an optical switch which re-routes the fibers after the first detection event (see Fig.~\ref{fig:QKD_schematic}).
This toggling prevents the node C from being idle and is a first step towards parallelization (see Sec.~\ref{sec:roadmap}).
This is one crucial technological step beyond the first demonstration of a memory-assisted QKD protocol in Ref.~\cite{Bhaskar2020}, where the authors employ a probabilistic routing of the photons.
In detail, the following switching rules are applied for this work:
\begin{enumerate}
    \item Initialize memory in state $\ket{\Psi}_\text{spin}^\text{init}\,,$ see Equation~(\ref{eq:spin-init}).
    \item Switch as soon as the successful loading of at least one memory is heralded.
    \item Perform BSM readout as soon as the successful loading of at least one memory, loaded already in step 2, is heralded.
    \item In case of successful BSM or maximal waiting time is reached, restart from step 1.
\end{enumerate}

The maximal waiting time in step 4 is introduced to upper bound the probability of losing a heralding photon before a successful BSM~\cite{Sohr2026_unpublished}, thereby limiting the noise in the loading process.

\begin{table}[h]
    \centering
    \caption{Outcomes of the BSM for matching bases. 
    Since the phases within each basis sum to $0$ or $\pm\pi$, the two possible BSM outcomes are $\mathrm{e}^{i(\phi+\phi')} = \pm1$. 
    Depending on the chosen basis and the BSM outcome, Alice or Bob flip their bit or leave it as it is. 
    After the BSM measurement is completed, both the basis choice and the BSM outcome are public information, while the bit values, encoded in the phase, are only private information of the communication parties Alice and Bob.
    Events with mismatched bases are discarded and therefore not displayed in this table.}
    \label{tab:BSM-outcomes}
    {
    \setlength{\arrayrulewidth}{0.5pt}
    \setlength{\tabcolsep}{0pt}
    \rowcolors{1}{white}{gray!30}
   \begin{tabularx}{0.6\columnwidth}{ 
   >{\hsize=0.3\hsize\centering\arraybackslash}X 
   >{\hsize=0.2\hsize\centering\arraybackslash}X 
   >{\hsize=0.2\hsize\centering\arraybackslash}X 
   >{\hsize=0.3\hsize\centering\arraybackslash}X}
    \hline
     Basis  & $\phi$   & $\phi'$  & $\mathrm{e}^{i(\phi+\phi')}$  \\
     \hline
      XX    & $0$      & $0$      &  $\phantom{-}1$\\
      XX    & $0$      & $\pi$    & $-1$\\
      XX    & $\pi$    & $0$      & $-1$\\
      XX    & $\pi$    & $\pi$    &  $\phantom{-}1$\\
      \hline
      YY    & $\phantom{-}\pi/2$  & $\phantom{-}\pi/2$  & $-1$\\
      YY    & $\phantom{-}\pi/2$  & $-\pi/2$ &  $\phantom{-}1$\\
      YY    & $-\pi/2$ & $\phantom{-}\pi/2$ &  $\phantom{-}1$\\
      YY    & $-\pi/2$ & $-\pi/2$ & $-1$\\
      \hline
    \end{tabularx}
    }
\end{table}
Provided with the BSM outcomes as shown in Table~\ref{tab:BSM-outcomes}, Alice or Bob flip their bit for the outcome $-1$ and $1$ if they used the $X$-basis and the $Y$-basis respectively. Otherwise they leave the key bit unchanged.

To get a benchmark for our simulation, we estimate the SKR $K$ in the asymptotic limit, assuming ideal single-photon sources and efficient encoding. The key rate depends on the source repetition rate $R$, the expected number of writing attempts per successful Bell-state measurement $N$, the system yield $Y$, and an error correction term characterized by the quantum bit error rate $e$,  the error correction efficiency $f$, and the binary Shannon entropy $h(e)$. This estimation follows the Devetak-Winter key rate formula~\cite{Devetak2005}:

\begin{equation}
    \label{eq:key-rate}
    K = \frac{R}{N} Y \left( 1 - \left[ \left( 1 + f \right) h(e) \right] \right)\,.
\end{equation}

The source repetition rate is given by \( R = 1/\tau_{\textrm{write}} \), where \( \tau_{\textrm{write}} \) is the time required to transfer a photonic state to the electron spin of the SiC device. 
The expected number of writing attempts per successful BSM, \( N \), accounts for the time required to load both Alice's and Bob’s states into memory and perform the BSM. 
This waiting time increases with transmission loss over distance but also depends on the chosen maximal waiting time.

The yield \( Y \) is the product of constant system characteristics, including the spin readout efficiency \( \eta_{\textrm{read}} \), the BSM efficiency \( \eta_{\textrm{BSM}} \), and the sifting ratio \( \eta_{\textrm{sift}} \), which reflects the fraction of retained measurement results after basis reconciliation.
The final factor in the SKR formula accounts for errors, where \( e \) is the quantum bit error rate and \( h(e) = -e\log_2e-(1-e)\log_2(1-e) \) is the binary Shannon entropy.

\subsection{Results and main limitations} \label{sec:results}
In this section, we analyze the key parameters of SiC devices and their impact on the performance of quantum nodes in MA-MDI-QKD. 
In long-distance QKD, the SKR is in general constrained by link loss, as well as by a combination of system parameters such as the photon rate, time resolution, and dark count rate of single-photon detectors~\cite{neumann2021model}.

For memory-assisted QKD to surpass direct point-to-point QKD in SKR, the advantages of quantum memories must outweigh the costs in efficiency, processing delays, and fidelity limitations of realistic SiC devices. 
Therefore, we focus on analyzing the impact of key parameters governing the performance of the asynchronous BSM, including efficiency, fidelity, communication latency, and processing time.

While we explicitly model imperfections such as finite detector quantum efficiency, dark counts, and insertion losses in optical switches and circulators, we assume ideal cavity cooperativity and spin readout fidelity for feasibility. 
Chiefly, we assume that the high cooperativity results in unity contrast for the two different spin states. Such approximations are suitable for the performance projections we aim for in this work.
A complete set of simulation parameters is provided in App.~\ref{sec:app}.  Detailed calculations can be performed to refine these estimates for given experimental parameters~\cite{Hanks2017}.

A fundamental limitation in the system arises from the writing process into the spin memory. 
While spin readout fidelity is high~\cite{Azuma2022}, ensuring accurate measurement of stored quantum states, spin-photon entanglement fidelity is significantly lower. 
This results from photon loss, decoherence, and the probabilistic nature of single-photon interference, which limits the efficiency of transferring a photonic qubit into the spin memory. 
The loss in the memory writing process not only reduces the SKR but additionally introduces noise during memory loading process, further degrading performance.

We show with our simulation how to reduce the additional loading noise and how to increase the heralding efficiency, allowing MA-MDI-QKD outperform both point-to-point QKD and standard MDI-QKD in terms of SKR.

Memory-assisted quantum networks rely on the ability to store quantum states with high fidelity over time.
In long-distance quantum links, low transmission probabilities necessitate storage to synchronize the arrival of two photons for the BSM.
A noteworthy limitation of MA-MDI-QKD therefore pertains the decay of the state fidelity of the stored state due to dephasing.
Dephasing is a loss of coherence due to interaction of the electron spin with the environment reducing BSM fidelity over time.
The smaller the coherence time, the faster the decay, resulting in higher the noise, lower the key rates.
While transferring the state from the electron spin to the long-lived nuclear spin could significantly extend the coherence time, practical limitations arise: conversion times fall within the microsecond regime, extending the writing time and thereby reducing the maximal possible source repetition rate, and nuclear spins do not couple to the cavity mode, making quantum registers impractical for this protocol.
In our system, dephasing occurs continuously from the last spin state initialization to the final spin readout that concludes the BSM, including both switching and waiting times.

\begin{figure}[t]
    \centering
    \includegraphics[width=1\linewidth]{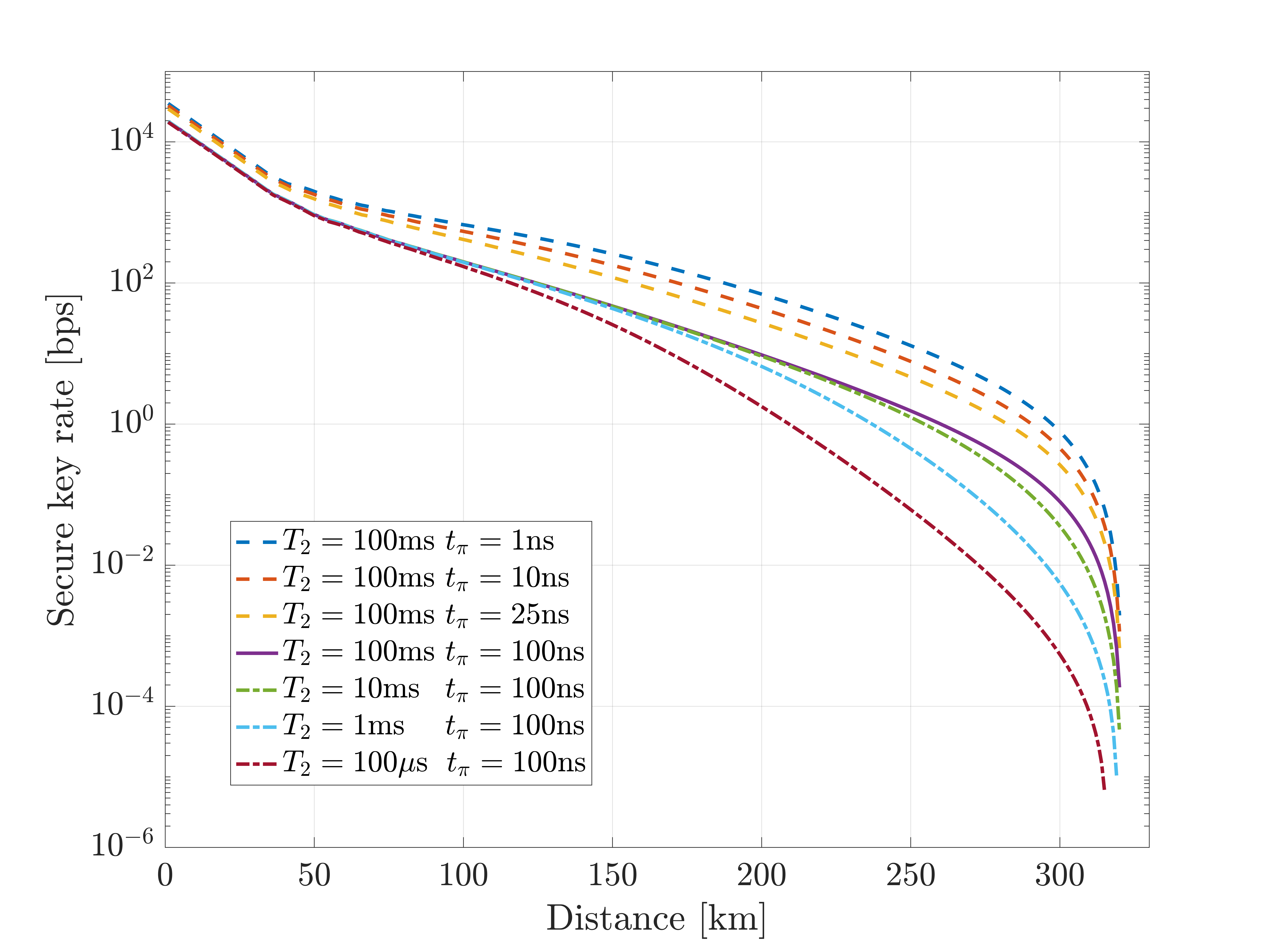}
    \caption{MA-MDI-QKD implemented with realistic devices and varying dephasing-time $T_2$ (dashed) and \(\pi\)-pulse durations $\tau_\pi$ (dash-dotted). 
    Secure key rates are plotted as a function of total separation $L$ between Alice and Bob. 
    Photons are transmitted over segments of $L/2$ each to the middle station via single-mode optical fiber (Corning SMF-28).
    The maximal waiting times are optimized for each set of parameters and each distance.
    For distances below 50 km the maximal waiting time is on the order of the inverse pulse rate, $1/R$.
    The spin is initialized after every possible time slot to write a qubit.
    The SKR scaling with the distance improves once waiting longer enables synchronising the signals from Alice and from Bob.
    Beyond 300 km the channel loss is so high, that the dark counts at 10 Hz dominate the signal-to-noise-ration causing a drop in the SKR to zero.
    The secure key rates are calculated in the asymptotic limit.}
    \label{fig:SKR_T2_tPi}
\end{figure}

The influence of the coherence time $T_2$ on the SKR can be observed in Fig.~\ref{fig:SKR_T2_tPi}. We simulated the SKR of MA-MDI-QKD for four different coherence times (\SI{100}{\micro\second}, \SI{1}{\milli\second}, \SI{10}{\milli\second}, and \SI{100}{\milli\second}), as a function of the link distance in standard telecommunication fibers (Corning SMF-28).
For link lengths below \SI{100}{\kilo\meter}, the dephasing has minimal impact on the SKR.
For distances between \SI{100}{\kilo\meter} and the maximal reach, where dark counts dominate, the effect of dephasing on the SKR becomes clear.
For shorter coherence times, dephasing reduces the key rate substantially.
Longer coherence times allow favorable scaling to be maintained over greater distances. 
Additionally, the sensitivity of the SKR to coherence time decreases as $T_2$ increases. 
The region where SKR scaling is dephasing-limited shifts toward longer transmission distances with increasing coherence time.
To maintain high key rates at long distances ($\gtrsim 100\,$km), coherence times of at least \SI{10}{\milli\second} are required.

Writing a photonic qubit into the quantum memory presents a time-consuming process when compared to the timescales of photon sources, which typically operate in the GHz regime. 
The primary constraint arises from the microwave pulse employed to perform unitary operations on the electron spin. 
The duration of a \(\pi\)-pulse, denoted as $\tau_\pi$, determines the minimum separation between the time-bins of a  qubit $\Delta t$, requiring $\Delta t \geq \tau_\pi$. 
Due to the probabilistic heralding measurement, photons are detected outside their initial temporal qubit space. 
Consequently, the time-bin separation of a single qubit also restricts the spacing between two successive time-bin qubits.
Thus, the photon repetition rate of the QKD transmitter is bounded by $1/(2\tau_\pi)$, where a typical \(\pi\)-pulse duration of $\tau_\pi=\SI{100}{\nano\second}$ yields a maximum photon repetition rate of 5 MHz. 
Furthermore, the photon repetition rate incurs an additional penalty resulting from the temporal spread of the time bins, which is constrained by the bandwidth of the quantum memory as determined by the time-bandwidth product of a transform-limited pulse. As an example, for $\tau_p = \SI{11.2}{\nano\second}$, the integral in a frequency window of 100 MHz exceeds 99.95$\,\%$.
Including this contribution, the photon repetition rate is ultimately bounded by $1/(2\tau_\pi+2\tau_p) = \SI{4.5}{\mega\hertz}$. For $t_\pi = \SI{10}{\nano\second}$, the repetition rate increases to $\SI{23.5}{\mega\hertz}$.

In Fig.~\ref{fig:SKR_T2_tPi}, the SKR is computed for the $\pi$-pulse duration taking the values (\SI{1}{\nano\second}, \SI{10}{\nano\second}, \SI{25}{\nano\second} and \SI{100}{\nano\second}). 
While the coherence time becomes relevant at long distances, the $\pi$-pulse length affects the SKR at every distance: A smaller $\pi$-pulse duration leads to a marked improvement for all distances, and has minor effects on the SKR scaling.

Despite this improvement, the SKR at \( L = \SI{0}{\kilo\meter} \) is nearly three orders of magnitude lower than the source repetition rate. 
This reduction arises from the impact of the expected number of writing attempts \( N \) in the key rate formula (Equation~\ref{eq:key-rate}). 
At short distances, the maximal waiting time is set to a small value to minimize loading errors. 
The trade-off is frequent and time-consuming memory reinitialization~\cite{Sohr2026_unpublished}, while the error term remains close to one, as loading errors are absent and dephasing effects are minimal.

\begin{figure}[t]
    \centering
    \includegraphics[width=1\linewidth]{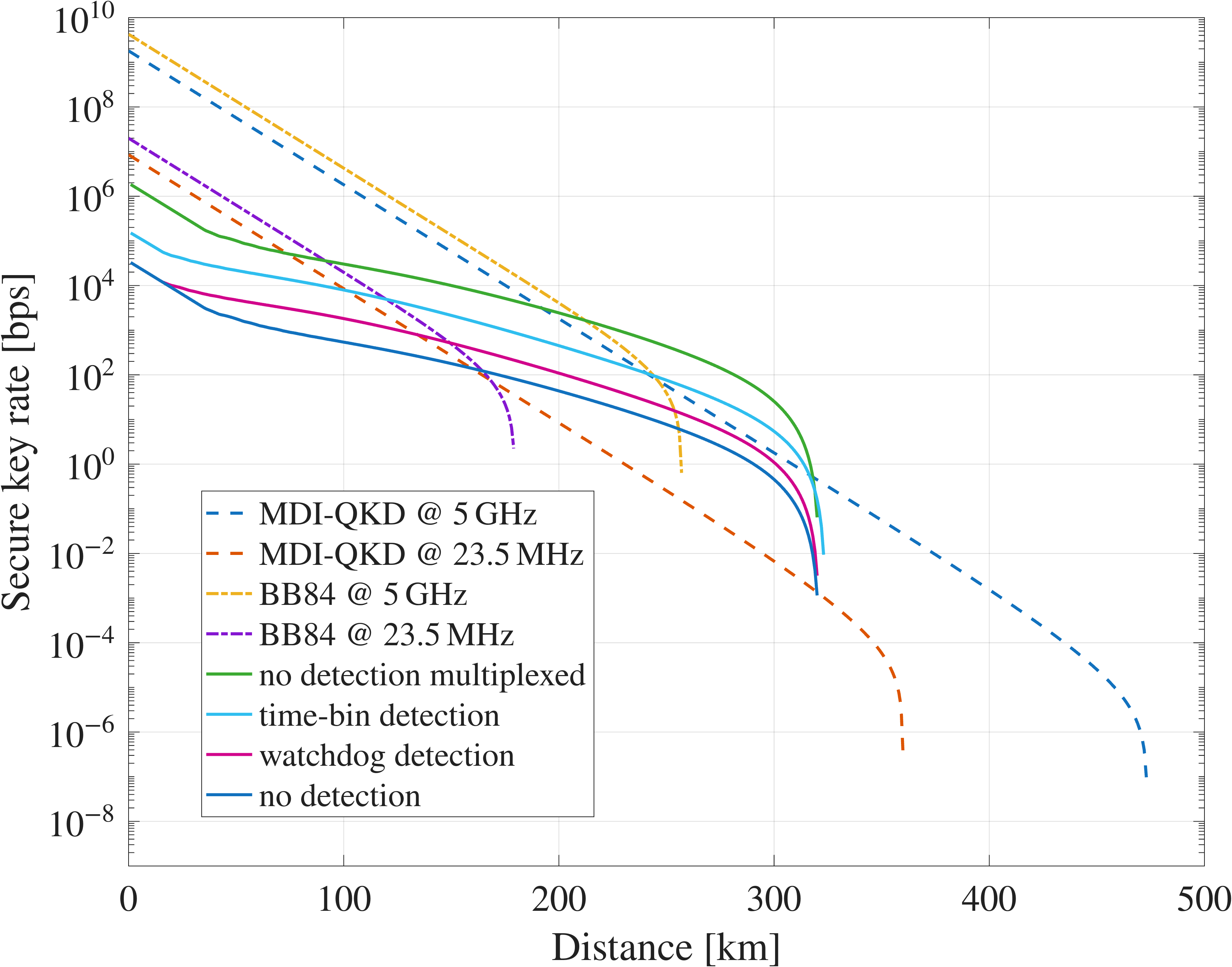}
    \caption{
    MA-MDI-QKD performance with different detection strategies in cavity transmission.
    Secure key rates are plotted as a function of total separation $L$ between Alice and Bob. 
    Photons are transmitted over segments of $L/2$ each to the middle station via single-mode optical fiber (Corning SMF-28).
    The maximal waiting time is optimized for maximum SKR at each distance, with coherence time and $\pi$-pulse duration fixed at $T_2=\SI{100}{\milli\second}$ and $t_\pi=\SI{10}{\nano\second}$. 
    The dark count rate is \SI{10}{\hertz}.
    As benchmarks, BB84 and MDI-QKD are shown at different photon repetition rates (\SI{23.5}{\mega\hertz} and \SI{5}{\giga\hertz}). 
    The SKR is calculated in the asymptotic limit. 
    For the multiplexed curve, the no-detection curve is scaled by a multiplexing factor of 56.
    All detection configurations in the cavity transmission outperform protocols operating at \SI{23.5}{\mega\hertz}. 
    However, only time-bin detection and the multiplexed no-detection setup achieve higher key rates than MDI-QKD at \SI{5}{\giga\hertz} for fiber lengths between \SI{250}{\kilo\meter} and \SI{310}{\kilo\meter}.
    }
    \label{fig:SKR_setup-choice-03dB}
\end{figure}

A further bottleneck of the scheme is the low heralding efficiency, constrained by an upper limit of \SI{25}{\percent}.
This sets a fundamental baseline for the efficiency of the asynchronous Bell state measurement in its current implementation.
A low heralding efficiency not only reduces the signal rate but also increases the noise, as the probability of undetected photon interaction with the cavity system increases.
We now analyze how monitoring photon-cavity interactions that do not lead to successful heralding increases the SKR.

Two types of non-heralding photon-cavity interactions can be monitored to reduce loading noise.
Firstly, the probabilistic time-bin detection has a success probability of \SI{50}{\percent}.
Instead of ignoring the unsuccessful events, they can be used to trigger immediate re-initialization of the memory.
This change can be implemented through an update in data collection and processing, without hardware modifications.
Secondly, photons transmitted rather than reflected by the cavity can be detected.

We investigate two detection strategies for the cavity transmission:
(i) A watchdog detector that registers transmission events without further analysis but triggers memory reinitialization upon detection.
As shown in Fig.~\ref{fig:SKR_setup-choice-03dB}, this shortens the linear-scaling region at short fiber lengths, shifting the SKR to higher values at larger distances.
(ii) Extending full time-bin detection, as used in the reflection path, to the transmission path of the cavity.
While experimentally demanding and potentially cost-intensive, this approach shifts the SKR to higher values across all fiber lengths, significantly enhancing system performance.

All configurations of transmission detection outperform BB84 and MDI-QKD when operated at the same repetition rate, benefiting from favorable scaling due to the asynchronous BSM.
With a watchdog detector, the SKR nearly surpasses that of MDI-QKD at a state-of-the-art repetition rate of \SI{5}{\giga\hertz}.
Employing a second time-bin detection module in the transmission of the cavity further boosts the key rate, enabling MA-MDI-QKD to outperform both direct transmission and MDI-QKD at \SI{5}{\giga\hertz}.

Independently of the detection strategy, wavelength multiplexing offers a path to higher SKR.
Multiplexing techniques, commonly used in modern telecommunication networks, can be harnessed to significantly enhance quantum network transmission rates. Wavelength multiplexing in fiber networks is highly standardized, with many cost-effective, off-the-shelf components available. 

The multiplexing factor of 56, which is used in Fig.~\ref{fig:SKR_setup-choice-03dB}, is derived by comparing the spectral width of the photons written to a SiC device with that of the 5 GHz system, assuming Fourier-limited linewidths.

Multiplexing not only offers higher key rates between two fixed communication parties, but also enables the extension to networks~\cite{wengerowsky2018entanglement} with several tens of users.
Using active optical switches~\cite{hall2021survey}, such a network could be extended to a star topology. 
Pairs of user nodes are then connected to a single SiC defect at the access node, where an asynchronous BSM is performed. 

Beyond the configurations we simulate, several modifications could further improve performance.
First, the non-interfering detection events in time-bin measurement could speed-up memory reinitialization.
These events do not herald successful writing, but they project the spin into a pure state, rendering classical readout unnecessary.
The initialization time thereby reduced to the duration of a $\pi/2$-pulse needed to restore the $\ket{+}$ state.
For the spin to be projected into a pure state, the detection events of two subsequent qubits must not overlap, reducing the rate by one third.
It remains to be investigated if the benefit of faster reinitialization outweighs the reduced rate due to qubit separation.

Second, the passive Mach-Zehnder interferometer used for heralding could be replaced with an actively switched version. 
Due to the probabilistic nature of the passive interferometer, half of the photons cannot be used to herald successful writing to the memory. 
Active switching, if achievable with sufficiently low insertion loss, would recover these photons and increase heralding efficiency.

Finally, it should be mentioned that photon losses could be further reduced by utilizing a different optical frequency.
The photon loss in the telecommunication O-band ($\sim\SI{0.3}{\decibel\per\kilo\meter}$), investigated here, is much lower than at optical wavelengths emitted by other defects~\cite{awschalom2018quantum}.
Still, the band with the lowest loss, which is primarily used in optical telecommunication systems, is the C-band, with a loss of $\sim\SI{0.2}{\decibel\per\kilo\meter}$.
While the difference between these two bands is seemingly small, it accumulates to an order of magnitude loss difference after \SI{100}{\kilo\meter} of propagation, providing strong motivation for the investigation of spin centres in this wavelength regime~\cite{Anisimov2025chlorine}.

\section{Roadmap for SiC quantum links} \label{sec:roadmap}

\begin{figure*}[t]
    \centering
    \includegraphics[width=1\textwidth]{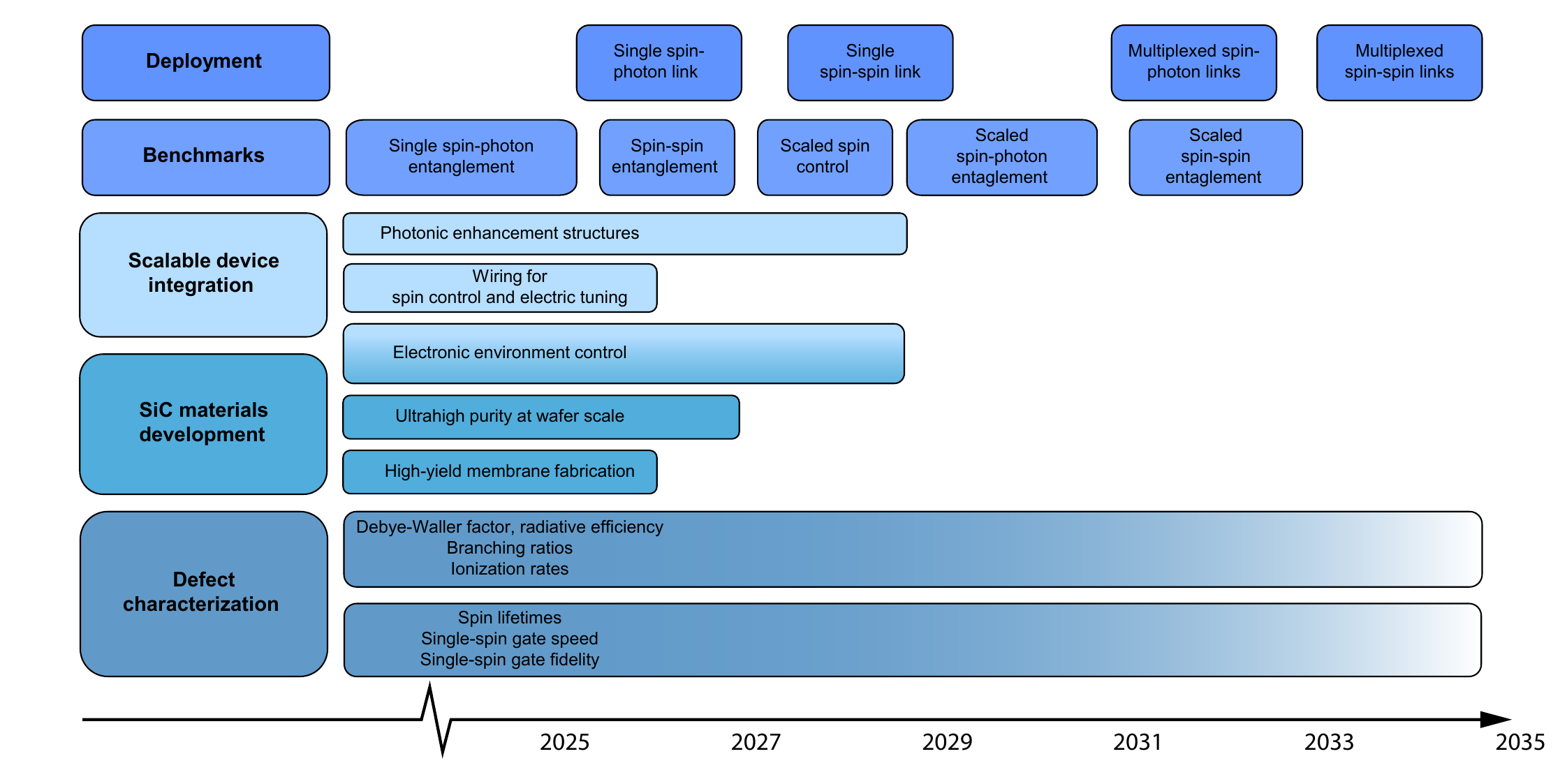}
    \caption{Roadmap for deployment of high-performance quantum links based on SiC photonics in the next decade. Bar ends indicate when technological developments (materials, integration) are expected to have reached a level suitable for large-scale deployment. Similarly, bar ends for benchmarks indicate when a performance level sufficient for deployment on metropolitan-scale links is achieved. Defect identification, characterization, and development is expected to lead to continued improvements throughout this evolution.}
    \label{fig:roadmap}
\end{figure*}

We now summarize the required technological steps for SiC photonics towards full deployment based on the example of repeater-extended quantum communication.  In order to be competitive with other approaches, this target needs to be reached within the next decade (see Fig. \ref{fig:roadmap}). As underlined by the preceding calculations, this application requires high-rate spin-photon entanglement with high fidelity and long ($\gg100\,$ms) spin coherence lifetime, with hundreds of addressable sites at each repeater node for multiplexing, to usefully exceed the rates provided by direct links.

\subsubsection{Prerequisites}
Regarding \textbf{\textit{materials development}}, the properties of most defects in SiC can be improved by carefully controlling the properties of the host crystal~\cite{son_APL2020}. Important aspects are the control of dopant density for charge state stability, a pristine crystalline structure to avoid strain inhomogeneity and undesired charge traps, and isotopic control to increase spin coherence as well as to reduce inhomogeneous broadening of the optical and spin transitions in some defects. Additionally, most if not all defects will benefit from charge depletion, which can be achieved in diode structures. A further necessity is the development of high-yield processes for the creation of thin, ultrasmooth membranes with uniform, $\SI{}{\micro\metre}$-scale thickness for FP microcavities and even thinner structures for PC cavities. While proof-of-principle demonstrations of such processes have also been performed, the yield or crystal quality of these SiC membranes have not yet reached a sufficient level for large-scale deployment. 

This development is necessary for all types of \textbf{\textit{optical enhancement structures}}~\cite{Castelletto2022}, apart from very few exceptions such as bullseye ring resonators, which can conceivably be implemented on bulk crystals but are not suitable for the spin-dependent reflection scheme discussed above. While defects which are spectrally and electronically stable might be utilized as emitters with Purcell or even only collection enhancement, the type of spin-photon interaction described above will be advantageous for most systems as it greatly reduces defect ionization and light-induced spectral diffusion.  We underline here that defects with longer wavelength transitions benefit from lower losses and a higher tolerance to imperfections in the photonic enhancement structures: As an example, the loss due to interface roughness in FP cavities decreases with $\sigma \propto 1/\lambda^2$, leading to a corresponding increase of the enhancement. Only very few examples of photonic enhancement have been reported for SiC defects so far, making the development of these structures an important task for the scientific community.

All \textbf{\textit{candidate defect centres}} will require high-fidelity spin initialization, manipulation, and readout, and will need to reach the desired coherence times. Additionally, high-yield creation of defects at predetermined positions in the crystal is highly desirable for the subsequent integration into photonic enhancement structures. In order to reduce the complexity of the multiplexing architecture, it is furthermore desirable to tune all defects at one repeater site to the same optical frequency so as to achieve indistinguishability. For most defects, this requirement implies tuning by strain or electrostatic means. Together with the requirement for fast individual spin control, this necessity implies a high degree of integration with multiple wire structures on the SiC surface, for which stringent limits on the power dissipation and conductive thermal load are given by the cooling power of the cryogenic environment. This, in turn, hints at the necessity for resonantly enhanced and/or superconducting spin control structures.

Many of the postulated requirements imply the use of \textbf{\textit{cryogenic infrastructure}} for all candidate defects, though only vanadium is known to require temperatures $< 1\,$K for sufficiently long spin relaxation lifetimes, while impressive coherence times have been achieved with other defect systems at around $4\,$K.

Finally, \textbf{\textit{wavelength conversion}} will be necessary for most of the known defects in SiC. The development of compact, scalable, and efficient converters will thus be necessary for most candidate defects. Successful proof-of-principle demonstrations have been achieved, but engineering a solution that is deployable at scale remains an outstanding task.

\subsubsection{Applications}
Photonic quantum links with memory units are broadly seen as having four distinct applications, which pose slightly different requirements:
    
\textbf{\textit{Repeaters for quantum key distribution}} that are integrated in a global network additionally require an efficient interface to space-based quantum communication systems, which currently operate at visible wavelengths or in the telecom C-band.   
    
\textbf{\textit{Photonically connected qubits or qudits within spin-based quantum computers}} can be envisioned without wavelength conversion or frequency multiplexing. Fault-tolerant architectures, however,  place even greater demands on yield, scalability, coherence, and fidelity of the SPI's. Additionally, they can benefit more strongly from nuclear spin ancillae than simple communication protocols. \CB{Quantum computing architectures based on small quantum processing units connected by a photonic channel~\cite{benjamin_PRX_2014}, all integrated on a single SiC chip including elements such as waveguides, Bragg filters, electrically-controlled optical modulators for routing, etc., can be envisioned.}

Further applications will be entanglement distribution for other \textbf{\textit{networked quantum computers}} and entanglement distribution for \textbf{\textit{networked quantum sensors}}.
For these, interfacing between the qubits or quantum sensors and the SPI needs to be developed.

\subsubsection{Prioritization}
These differences notwithstanding, the overall set of requirements allows to set priority levels for development of different aspects of the technology. We focus here on aspects which are specifically connected to SiC devices, thereby excluding wavelength conversion and cryogenic infrastructure, to which advances in SiC photonics cannot contribute directly.
The highest priority should be given to materials development aspects, particularly the development of wafer-scale membranes. Most of these are additionally beneficial to the SiC industry and to emerging applications of classical SiC photonics, thereby providing the highest incentives overall. The next priority is the selection of the most suitable defect for deployment, as this determines the requirements for photonic enhancement and interoperability with telecom networks. As several defects currently under study present advantageous features, this decision can only be taken at a later stage of technological development. Nonetheless, several performance parameters need to be ascertained precisely along this path, including the Debye-Waller factor, radiative efficiency, and transition branching ratios, the spin relaxation and coherence lifetimes, \CB{the operation temperature}, as well as any limitations on the single-spin initialization, readout, as well as gate times and fidelities which could limit the SPI performance. This type of characterization can be expected to exclude several defects from applications in quantum photonics. However, as new candidate defects continue to emerge, this task will remain ongoing beyond the deployment of the first devices.

For SPI benchmarking, each defect with sufficient performance parameters needs to be incorporated in a photonic enhancement structure, replete with all control wiring and potentially with SiC diode structures. Aside from the enhancement factor, the main emphasis in this development must be placed on scalability, as vast numbers of SPI's will be required for all quantum technology applications. A large part of these technological aspects is common to all defects and should therefore be performed in parallel to the detailed characterization of spin centres.

Since different applications and protocols place differing requirements on the performance of the SPI, a global benchmark cannot be defined. Nonetheless, certain performance indicators are common to most settings. A possible baseline benchmark which encapsulates many of the stated requirements is the product of the spin-photon entanglement rate and the spin coherence lifetime at telecom wavelength and after fiber coupling. The next benchmark level is the product of rate and lifetime of spin-spin entanglement with two such units, which subsumes aspects such as indistinguishability and synchronization. Both of these benchmarks need to be defined with a threshold fidelity, or be scaled appropriately for a chosen protocol.

\subsubsection{Timeline}
 For these developments to take place within the next decade, a concerted effort by the SiC photonics community with strong industrial participation will be required. The payoff, however, will be large for all involved stakeholders, and will provide technological developments with beneficial effects far beyond the science and technology domains. While ambitious, the timeline for these developments can be achieved assuming continued growth in support from governments and continued advances in the burgeoning SiC industry. 

\section{Summary}
\label{sec:summary}

We have provided an overview of SiC photonics using spin centres for quantum links. These devices have applications in quantum computing, secure key distribution, and as entanglement links for distributed quantum computation and sensing. Using memory-assisted QKD as an example, we have investigated the key parameters which determine the performance of such links.

We have undertaken a comprehensive analysis of the key parameters of SiC devices and their impact on the performance of quantum nodes in the context of memory-assisted measurement-device-independent-QKD. 
We opted for this particular QKD protocol as it provides a practical benchmark for examining the performance of a single quantum memory by focusing on a single parameter, the secure key rate.
Importantly, the implementation of this protocol with SiC devices includes all essential building blocks of large-scale repeater networks. 
Our investigation has revealed several crucial factors that influence the secure key rate over long communication distances, shedding light on both the potential and limitations of spin centers in SiC.
Furthermore, the SiC platform's compatibility with wavelength multiplexing enables scaling to multi-user networks, as discussed in Section~\ref{sec:results}.

Quantum repeater networks are the ultimate goal for long-distance quantum communication and the SiC platform is a strong candidate for serving this purpose. 
This vision can only be realized once a single quantum repeater node outperforms point-to-point quantum communication. 
Only then, quantum repeater networks with several quantum repeater nodes will prove beneficial in quantum communication and might be deployed in the field. 
This should also guide the experimental efforts in SiC-based SPI's. 
Therefore, the first milestone on the roadmap towards a large-scale quantum repeater network is the demonstration of the single-node~\cite{Bhaskar2020} or two-node~\cite{Liu2021,Lago2021} scenario.

Beyond beating the repeaterless bound~\cite{pirandola2017fundamental} per channel use~\cite{Rozpdek2019, Bhaskar2020}, we benchmark SiC devices with existing QKD technologies, which, for example, allow higher photon repetition rates compared to current specifications of CQED devices.
This practical approach allows us to make statements about the operational capability of SiC-based quantum memories in real-world quantum networks. In particular, the results highlight the need for multiplexing in order to achieve competitive secret key rates. Scalability in SPI production and interconnects is therefore a requirement for any quantum repeater technology based on spin-photon interfaces.

Another usecase is the deployment of SiC devices for end nodes in quantum networks.
The quantum memories in end nodes delay the read-out of photonic quantum states or re-emit stored states after certain time intervals for various quantum information processing tasks.
End nodes therefore have a different set of requirements compared to repeater nodes.   
As an example, the transfer to nuclear spins is a good option to increase the storage time of end nodes, even though in-depth comparison to other quantum memory platforms is necessary to evaluate the deployability of defects in SiC for this usecase. 

These considerations allow to define a path towards deployment of SiC photonics for high-performance quantum links, outlined in Sec.\ref{sec:roadmap}. We project that, with a concerted effort of the scientific community and the SiC industry, it will be possible to achieve this goal within a decade.

\section{Acknowledgements}
This work was funded by the European Union under grant agreements No 862721 (QuanTELCO) and 101186889 (QuSPARC). We acknowledge support from the Austrian Research Promotion Agency project under projects FFG FO999914034 (SPQV) and FO999921415 (VANESSA\_QC).

\bibliography{references}

@article{Pettersson2025,
  title={Long-distance quantum communication using concatenated ring graph codes},
  author={Pettersson, Love and S{\o}rensen, Anders S},
  journal={arXiv preprint arXiv:2503.19822},
  year={2025}
}

@article{Wo2023,
  title={Resource-efficient fault-tolerant one-way quantum repeater with code concatenation},
  author={Wo, Kah Jen and Avis, Guus and Rozp{\k{e}}dek, Filip and Mor-Ruiz, Maria Flors and Pieplow, Gregor and Schr{\"o}der, Tim and Jiang, Liang and S{\o}rensen, Anders S and Borregaard, Johannes},
  journal={npj Quantum Information},
  volume={9},
  number={1},
  pages={123},
  year={2023},
  publisher={Nature Publishing Group UK London}
}

@article{Muralidharan2016,
  title={Optimal architectures for long distance quantum communication},
  author={Muralidharan, Sreraman and Li, Linshu and Kim, Jungsang and L{\"u}tkenhaus, Norbert and Lukin, Mikhail D and Jiang, Liang},
  journal={Scientific reports},
  volume={6},
  number={1},
  pages={20463},
  year={2016},
  publisher={Nature Publishing Group UK London}
}

@article{Hanks2017,
  title={High-fidelity spin measurement on the nitrogen-vacancy center},
  author={Hanks, Michael and Trupke, Michael and Schmiedmayer, J{\"o}rg and Munro, William J and Nemoto, Kae},
  journal={New Journal of Physics},
  volume={19},
  number={10},
  pages={103002},
  year={2017},
  publisher={IOP Publishing}
}

@article{hessenauer2025cavityenhancementv2centers,
  title={Cavity enhancement of V2 centers in 4H-SiC with a fiber-based Fabry--Perot microcavity},
  author={Hessenauer, Jannis and K{\"o}rber, Jonathan and Ghezellou, Misagh and Ul-Hassan, Jawad and Astakhov, Georgy V and Knolle, Wolfgang and Wrachtrup, J{\"o}rg and Hunger, David},
  journal={Optica Quantum},
  volume={3},
  number={2},
  pages={175--181},
  year={2025},
  publisher={Optica Publishing Group}
}

@misc{koller2025strainenabledcontrolvanadiumqudit,
      title={Strain-enabled control of the vanadium qudit in silicon carbide}, 
      author={Philipp Koller and Thomas Astner and Benedikt Tissot and Guido Burkard and Michael Trupke},
      year={2025},
      eprint={2501.05896},
      archivePrefix={arXiv},
      primaryClass={quant-ph},
      url={https://arxiv.org/abs/2501.05896}, 
}

@Article{Horodecki2009a,
  author           = {Horodecki, Ryszard and Horodecki, Paweł and Horodecki, Michał and Horodecki, Karol},
  journal          = {Reviews of Modern Physics},
  title            = {Quantum entanglement},
  year             = {2009},
  month            = jun,
  number           = {2},
  pages            = {865--942},
  volume           = {81},
  doi              = {10.1103/RevModPhys.81.865},
  publisher        = {American Physical Society},
}

@Article{Azuma2023,
  author           = {Azuma, Koji and Economou, Sophia E and Elkouss, David and Hilaire, Paul and Jiang, Liang and Lo, Hoi-Kwong and Tzitrin, Ilan},
  journal          = {Reviews of Modern Physics},
  title            = {Quantum repeaters: From quantum networks to the quantum internet},
  year             = {2023},
  month            = dec,
  number           = {4},
  pages            = {045006},
  volume           = {95},
  doi              = {10.1103/RevModPhys.95.045006},
  publisher        = {APS},
  shorttitle       = {Quantum repeaters},
}

@Article{Devetak2005,
  author           = {Devetak, Igor and Winter, Andreas},
  journal          = {Proceedings of the Royal Society A: Mathematical, Physical and Engineering Sciences},
  title            = {Distillation of Secret Key and Entanglement from Quantum States},
  year             = {2005},
  month            = jan,
  number           = {2053},
  pages            = {207--235},
  volume           = {461},
  doi              = {10.1098/rspa.2004.1372},
  publisher        = {{Royal Society}},
}

@misc{Sohr2026_unpublished,
  author = {Sohr, Philipp and Koller, Philipp and Astner, Thomas and Trupke, Michael},
  year = {2026},
  note = {Manuscript in preparation.}
}

@article{son_APL2020,
    author = {Son, Nguyen T. and Anderson, Christopher P. and Bourassa, Alexandre and Miao, Kevin C. and Babin, Charles and Widmann, Matthias and Niethammer, Matthias and Ul Hassan, Jawad and Morioka, Naoya and Ivanov, Ivan G. and Kaiser, Florian and Wrachtrup, Joerg and Awschalom, David D.},
    title = "{Developing silicon carbide for quantum spintronics}",
    journal = {Applied Physics Letters},
    volume = {116},
    number = {19},
    pages = {190501},
    year = {2020},
    month = {05},
    issn = {0003-6951},
    doi = {10.1063/5.0004454},
    url = {https://doi.org/10.1063/5.0004454},
    eprint = {https://pubs.aip.org/aip/apl/article-pdf/doi/10.1063/5.0004454/14532891/190501\_1\_online.pdf},
}

@article{benjamin_PRX_2014,
  title = {Freely Scalable Quantum Technologies Using Cells of 5-to-50 Qubits with Very Lossy and Noisy Photonic Links},
  author = {Nickerson, Naomi H. and Fitzsimons, Joseph F. and Benjamin, Simon C.},
  journal = {Phys. Rev. X},
  volume = {4},
  issue = {4},
  pages = {041041},
  numpages = {17},
  year = {2014},
  month = {Dec},
  publisher = {American Physical Society},
  doi = {10.1103/PhysRevX.4.041041},
  url = {https://link.aps.org/doi/10.1103/PhysRevX.4.041041}
}

@inproceedings{10.1117/12.2689972,
author = {Sebastian Ecker and Johannes Pseiner and Jorge Piris and Martin Bohmann},
title = {{Advances in entanglement-based QKD for space applications}},
volume = {12777},
booktitle = {International Conference on Space Optics — ICSO 2022},
editor = {Kyriaki Minoglou and Nikos Karafolas and Bruno Cugny},
organization = {International Society for Optics and Photonics},
publisher = {SPIE},
pages = {1277727},
year = {2023},
doi = {10.1117/12.2689972},
URL = {https://doi.org/10.1117/12.2689972}
}

@article{bracher_selective_2017,
   author = {David O. Bracher and Xingyu Zhang and Evelyn L. Hu},
   doi = {10.1073/pnas.1704219114},
   issn = {0027-8424},
   issue = {16},
   journal = {Proceedings of the National Academy of Sciences},
   month = {4},
   pages = {4060-4065},
   title = {Selective Purcell enhancement of two closely linked zero-phonon transitions of a silicon carbide color center},
   volume = {114},
   year = {2017},
}

@article{lukin_two-emitter_2023,
	title = {Two-{Emitter} {Multimode} {Cavity} {Quantum} {Electrodynamics} in {Thin}-{Film} {Silicon} {Carbide} {Photonics}},
	volume = {13},
	url = {https://link.aps.org/doi/10.1103/PhysRevX.13.011005},
	doi = {10.1103/PhysRevX.13.011005},
	number = {1},
	urldate = {2023-12-21},
	journal = {Physical Review X},
	author = {Lukin, Daniil M. and Guidry, Melissa A. and Yang, Joshua and Ghezellou, Misagh and Deb Mishra, Sattwik and Abe, Hiroshi and Ohshima, Takeshi and Ul-Hassan, Jawad and Vučković, Jelena},
	month = jan,
	year = {2023},
	pages = {011005},
}

@article{yang_inverse-designed_2023,
   author = {Joshua Yang and Melissa A. Guidry and Daniil M. Lukin and Kiyoul Yang and Jelena Vučković},
   doi = {10.1038/s41377-023-01253-9},
   issn = {2047-7538},
   issue = {1},
   journal = {Light: Science I\& Applications},
   month = {8},
   pages = {201},
   title = {Inverse-designed silicon carbide quantum and nonlinear photonics},
   volume = {12},
   year = {2023},
}

@article{bersin2023telecom,
  title={Telecom networking with a diamond quantum memory},
  author={Bersin, Eric and Sutula, Madison and Huan, Yan Qi and Suleymanzade, Aziza and Assumpcao, Daniel R and Wei, Yan-Cheng and Stas, Pieter-Jan and Knaut, Can M and Knall, Erik N and Langrock, Carsten and others},
  journal={arXiv preprint arXiv:2307.08619},
  year={2023}
}

@article{divincenzo2000physical,
  title={The physical implementation of quantum computation},
  author={DiVincenzo, David P},
  journal={Fortschritte der Physik: Progress of Physics},
  volume={48},
  number={9-11},
  pages={771--783},
  year={2000},
  publisher={Wiley Online Library}
}

@article{park1970concept,
  title={The concept of transition in quantum mechanics},
  author={Park, James L},
  journal={Foundations of physics},
  volume={1},
  pages={23--33},
  year={1970},
  publisher={Springer}
}

@article{wootters1982single,
  title={A single quantum cannot be cloned},
  author={Wootters, William K and Zurek, Wojciech H},
  journal={Nature},
  volume={299},
  number={5886},
  pages={802--803},
  year={1982},
  publisher={Nature Publishing Group UK London}
}

@article{kimoto2022high,
  title={High-voltage SiC power devices for improved energy efficiency},
  author={Kimoto, Tsunenobu},
  journal={Proceedings of the Japan Academy, Series B},
  volume={98},
  number={4},
  pages={161--189},
  year={2022},
  publisher={The Japan Academy}
}

@article{she2017review,
  title={Review of silicon carbide power devices and their applications},
  author={She, Xu and Huang, Alex Q and Lucia, Oscar and Ozpineci, Burak},
  journal={IEEE Transactions on Industrial Electronics},
  volume={64},
  number={10},
  pages={8193--8205},
  year={2017},
  publisher={IEEE}
}

@article{Taminiau12,
  title = {Detection and Control of Individual Nuclear Spins Using a Weakly Coupled Electron Spin},
  author = {Taminiau, T. H. and Wagenaar, J. J. T. and van der Sar, T. and Jelezko, F. and Dobrovitski, V. V. and Hanson, R.},
  journal = {Phys. Rev. Lett.},
  volume = {109},
  issue = {13},
  pages = {137602},
  numpages = {5},
  year = {2012},
  month = {Sep},
  publisher = {American Physical Society},
  doi = {10.1103/PhysRevLett.109.137602},
  url = {https://link.aps.org/doi/10.1103/PhysRevLett.109.137602}
}

@article{Zheng2019,
  doi = {10.1364/ol.44.005784},
  url = {https://doi.org/10.1364/ol.44.005784},
  year = {2019},
  month = nov,
  publisher = {The Optical Society},
  volume = {44},
  number = {23},
  pages = {5784},
  author = {Yi Zheng and Minhao Pu and Ailun Yi and Xin Ou and Haiyan Ou},
  title = {4H-{SiC} microring resonators for nonlinear integrated photonics},
  journal = {Optics Letters}
}

@article{Yi2020,
  doi = {10.1016/j.optmat.2020.109990},
  url = {https://doi.org/10.1016/j.optmat.2020.109990},
  year = {2020},
  month = sep,
  publisher = {Elsevier {BV}},
  volume = {107},
  pages = {109990},
  author = {Ailun Yi and Yi Zheng and Hao Huang and Jiajie Lin and Youquan Yan and Tiangui You and Kai Huang and Shibin Zhang and Chen Shen and Min Zhou and Wei Huang and Jiaxiang Zhang and Shengqiang Zhou and Haiyan Ou and Xin Ou},
  title = {Wafer-scale 4H-silicon carbide-on-insulator (4H{\textendash}{SiCOI}) platform for nonlinear integrated optical devices},
  journal = {Optical Materials}
}

@article{Shi2021,
  doi = {10.1063/5.0053296},
  url = {https://doi.org/10.1063/5.0053296},
  year = {2021},
  month = jul,
  publisher = {{AIP} Publishing},
  volume = {6},
  number = {7},
  author = {Xiaodong Shi and Weichen Fan and Yaoqin Lu and Anders Kragh Hansen and Mingjun Chi and Ailun Yi and Xin Ou and Karsten Rottwitt and Haiyan Ou},
  title = {Polarization and spatial mode dependent four-wave mixing in a 4H-silicon carbide microring resonator},
  journal = {{APL} Photonics}
}

@article{Ou2023,
  doi = {10.3390/ma16031014},
  url = {https://doi.org/10.3390/ma16031014},
  year = {2023},
  month = jan,
  publisher = {{MDPI} {AG}},
  volume = {16},
  number = {3},
  pages = {1014},
  author = {Haiyan Ou and Xiaodong Shi and Yaoqin Lu and Manuel Kollmuss and Johannes Steiner and Vincent Tabouret and Mikael Syv\"{a}j\"{a}rvi and Peter Wellmann and Didier Chaussende},
  title = {Novel Photonic Applications of Silicon Carbide},
  journal = {Materials}
}

@article{Shi2023,
  doi = {10.1364/ol.481314},
  url = {https://doi.org/10.1364/ol.481314},
  year = {2023},
  month = jan,
  publisher = {Optica Publishing Group},
  volume = {48},
  number = {3},
  pages = {616},
  author = {Xiaodong Shi and Yaoqin Lu and Haiyan Ou},
  title = {High-performance silicon carbide polarization beam splitting based on an asymmetric directional couplers for mode conversion},
  journal = {Optics Letters}
}

@article{LaVia2023,
  doi = {10.3390/mi14061200},
  url = {https://doi.org/10.3390/mi14061200},
  year = {2023},
  month = jun,
  publisher = {{MDPI} {AG}},
  volume = {14},
  number = {6},
  pages = {1200},
  author = {Francesco La Via and Daniel Alquier and Filippo Giannazzo and Tsunenobu Kimoto and Philip Neudeck and Haiyan Ou and Alberto Roncaglia and Stephen E. Saddow and Salvatore Tudisco},
  title = {Emerging {SiC} Applications beyond Power Electronic Devices},
  journal = {Micromachines}
}

@article{nguyen2019quantum,
  title={Quantum network nodes based on diamond qubits with an efficient nanophotonic interface},
  author={Nguyen, CT and Sukachev, DD and Bhaskar, MK and Machielse, Bartholomeus and Levonian, DS and Knall, EN and Stroganov, Pavel and Riedinger, Ralf and Park, Hongkun and Lon{\v{c}}ar, M and others},
  journal={Physical review letters},
  volume={123},
  number={18},
  pages={183602},
  year={2019},
  publisher={APS}
}

@article{awschalom2018quantum,
  title={Quantum technologies with optically interfaced solid-state spins},
  author={Awschalom, David D and Hanson, Ronald and Wrachtrup, J{\"o}rg and Zhou, Brian B},
  journal={Nature Photonics},
  volume={12},
  number={9},
  pages={516--527},
  year={2018},
  publisher={Nature Publishing Group UK London}
}

@article{neumann2021model,
  title={Model for optimizing quantum key distribution with continuous-wave pumped entangled-photon sources},
  author={Neumann, Sebastian Philipp and Scheidl, Thomas and Selimovic, Mirela and Pivoluska, Matej and Liu, Bo and Bohmann, Martin and Ursin, Rupert},
  journal={Physical Review A},
  volume={104},
  number={2},
  pages={022406},
  year={2021},
  publisher={APS}
}

@article{bennett1996purification,
  title={Purification of noisy entanglement and faithful teleportation via noisy channels},
  author={Bennett, Charles H and Brassard, Gilles and Popescu, Sandu and Schumacher, Benjamin and Smolin, John A and Wootters, William K},
  journal={Physical review letters},
  volume={76},
  number={5},
  pages={722},
  year={1996},
  publisher={APS}
}

@article{neumann2022continuous,
  title={Continuous entanglement distribution over a transnational 248 km fiber link},
  author={Neumann, Sebastian Philipp and Buchner, Alexander and Bulla, Lukas and Bohmann, Martin and Ursin, Rupert},
  journal={Nature Communications},
  volume={13},
  number={1},
  pages={6134},
  year={2022},
  publisher={Nature Publishing Group UK London}
}

@article{bennett1992quantum,
  title={Quantum cryptography without Bell’s theorem},
  author={Bennett, Charles H and Brassard, Gilles and Mermin, N David},
  journal={Physical review letters},
  volume={68},
  number={5},
  pages={557},
  year={1992},
  publisher={APS}
}

@article{bennett2014quantum,
  title={Quantum cryptography: Public key distribution and coin tossing},
  author={Bennett, Charles H and Brassard, Gilles},
  journal={Theoretical computer science},
  volume={560},
  pages={7--11},
  year={2014},
  publisher={Elsevier}
}

@article{Hammerer2010,
  title = {Quantum interface between light and atomic ensembles},
  author = {Hammerer, Klemens and S\o{}rensen, Anders S. and Polzik, Eugene S.},
  journal = {Rev. Mod. Phys.},
  volume = {82},
  issue = {2},
  pages = {1041--1093},
  numpages = {0},
  year = {2010},
  month = {Apr},
  publisher = {American Physical Society},
  doi = {10.1103/RevModPhys.82.1041},
  url = {https://link.aps.org/doi/10.1103/RevModPhys.82.1041}
}

@article{Lodahl2017,
  doi = {10.1088/2058-9565/aa91bb},
  url = {https://doi.org/10.1088/2058-9565/aa91bb},
  year = {2017},
  month = oct,
  publisher = {{IOP} Publishing},
  volume = {3},
  number = {1},
  pages = {013001},
  author = {Peter Lodahl},
  title = {Quantum-dot based photonic quantum networks},
  journal = {Quantum Science and Technology}
}

@article{quantumdot2010,
  title = {Quantum-Dot-Spin Single-Photon Interface},
  author = {Y\ifmmode \imath \else \i \fi{}lmaz, S. T. and Fallahi, P. and Imamo\ifmmode \breve{g}\else \u{g}\fi{}lu, A.},
  journal = {Phys. Rev. Lett.},
  volume = {105},
  issue = {3},
  pages = {033601},
  numpages = {4},
  year = {2010},
  month = {Jul},
  publisher = {American Physical Society},
  doi = {10.1103/PhysRevLett.105.033601},
  url = {https://link.aps.org/doi/10.1103/PhysRevLett.105.033601}
}

@article{Krutyanskiy2023,
  title = {Entanglement of Trapped-Ion Qubits Separated by 230 Meters},
  author = {Krutyanskiy, V. and Galli, M. and Krcmarsky, V. and Baier, S. and Fioretto, D. A. and Pu, Y. and Mazloom, A. and Sekatski, P. and Canteri, M. and Teller, M. and Schupp, J. and Bate, J. and Meraner, M. and Sangouard, N. and Lanyon, B. P. and Northup, T. E.},
  journal = {Phys. Rev. Lett.},
  volume = {130},
  issue = {5},
  pages = {050803},
  numpages = {7},
  year = {2023},
  month = {Feb},
  publisher = {American Physical Society},
  doi = {10.1103/PhysRevLett.130.050803},
  url = {https://link.aps.org/doi/10.1103/PhysRevLett.130.050803}
}

@article{Liao2018,
  title = {Satellite-Relayed Intercontinental Quantum Network},
  author = {Liao, Sheng-Kai and Cai, Wen-Qi and Handsteiner, Johannes and Liu, Bo and Yin, Juan and Zhang, Liang and Rauch, Dominik and Fink, Matthias and Ren, Ji-Gang and Liu, Wei-Yue and Li, Yang and Shen, Qi and Cao, Yuan and Li, Feng-Zhi and Wang, Jian-Feng and Huang, Yong-Mei and Deng, Lei and Xi, Tao and Ma, Lu and Hu, Tai and Li, Li and Liu, Nai-Le and Koidl, Franz and Wang, Peiyuan and Chen, Yu-Ao and Wang, Xiang-Bin and Steindorfer, Michael and Kirchner, Georg and Lu, Chao-Yang and Shu, Rong and Ursin, Rupert and Scheidl, Thomas and Peng, Cheng-Zhi and Wang, Jian-Yu and Zeilinger, Anton and Pan, Jian-Wei},
  journal = {Phys. Rev. Lett.},
  volume = {120},
  issue = {3},
  pages = {030501},
  numpages = {4},
  year = {2018},
  month = {Jan},
  publisher = {American Physical Society},
  doi = {10.1103/PhysRevLett.120.030501},
  url = {https://link.aps.org/doi/10.1103/PhysRevLett.120.030501}
}

@article{Lu2022,
  title = {Micius quantum experiments in space},
  author = {Lu, Chao-Yang and Cao, Yuan and Peng, Cheng-Zhi and Pan, Jian-Wei},
  journal = {Rev. Mod. Phys.},
  volume = {94},
  issue = {3},
  pages = {035001},
  numpages = {46},
  year = {2022},
  publisher = {American Physical Society},
  doi = {10.1103/RevModPhys.94.035001},
  url = {https://link.aps.org/doi/10.1103/RevModPhys.94.035001}
}

@article{Zhang2021,
  doi = {10.1088/2058-9565/abd4c3},
  url = {https://doi.org/10.1088/2058-9565/abd4c3},
  year = {2021},
  publisher = {{IOP} Publishing},
  volume = {6},
  number = {4},
  pages = {043001},
  author = {Zheshen Zhang and Quntao Zhuang},
  title = {Distributed quantum sensing},
  journal = {Quantum Science and Technology}
}

@article{Lo2012,
  title = {Measurement-Device-Independent Quantum Key Distribution},
  author = {Lo, Hoi-Kwong and Curty, Marcos and Qi, Bing},
  journal = {Phys. Rev. Lett.},
  volume = {108},
  issue = {13},
  pages = {130503},
  numpages = {5},
  year = {2012},
  publisher = {American Physical Society},
  doi = {10.1103/PhysRevLett.108.130503},
  url = {https://link.aps.org/doi/10.1103/PhysRevLett.108.130503}
}

@article{Franson1989,
   author = {J. D. Franson},
   doi = {10.1103/PhysRevLett.62.2205},
   isbn = {0031-9007},
   issn = {00319007},
   issue = {19},
   journal = {Physical Review Letters},
   pages = {2205-2208},
   pmid = {10039885},
   title = {Bell inequality for position and time},
   volume = {62},
   year = {1989},
}

@article{Tchebotareva2019,
  doi = {10.1103/physrevlett.123.063601},
  url = {https://doi.org/10.1103/physrevlett.123.063601},
  year = {2019},
  publisher = {American Physical Society ({APS})},
  volume = {123},
  number = {6},
  author = {Anna Tchebotareva and Sophie L.{\hspace{0.167em}}N. Hermans and Peter C. Humphreys and Dirk Voigt and Peter J. Harmsma and Lun K. Cheng and Ad L. Verlaan and Niels Dijkhuizen and Wim de Jong and Anaïs Dr{\'{e}}au and Ronald Hanson},
  title = {Entanglement between a Diamond Spin Qubit and a Photonic Time-Bin Qubit at Telecom Wavelength},
  journal = {Physical Review Letters}
}

@article{Rozpdek2019,
  doi = {10.1103/physreva.99.052330},
  url = {https://doi.org/10.1103/physreva.99.052330},
  year = {2019},
  month = may,
  publisher = {American Physical Society ({APS})},
  volume = {99},
  number = {5},
  author = {Filip Rozp\k{e}dek and Raja Yehia and Kenneth Goodenough and Maximilian Ruf and Peter C. Humphreys and Ronald Hanson and Stephanie Wehner and David Elkouss},
  title = {Near-term quantum-repeater experiments with nitrogen-vacancy centers: Overcoming the limitations of direct transmission},
  journal = {Physical Review A}
}

@article{Wehner2018,
   author = {Stephanie Wehner and David Elkouss and Ronald Hanson},
   doi = {10.1126/science.aam9288},
   issn = {0036-8075},
   issue = {6412},
   journal = {Science},
   pages = {eaam9288},
   title = {Quantum internet: A vision for the road ahead},
   volume = {362},
   url = {http://www.sciencemag.org/lookup/doi/10.1126/science.aam9288},
   year = {2018},
}

@article{Azuma2022,
  title = {Quantum repeaters: From quantum networks to the quantum internet},
  author = {Azuma, Koji and Economou, Sophia E. and Elkouss, David and Hilaire, Paul and Jiang, Liang and Lo, Hoi-Kwong and Tzitrin, Ilan},
  journal = {Rev. Mod. Phys.},
  volume = {95},
  issue = {4},
  pages = {045006},
  numpages = {66},
  year = {2023},
  month = {Dec},
  publisher = {American Physical Society},
  doi = {10.1103/RevModPhys.95.045006},
  url = {https://link.aps.org/doi/10.1103/RevModPhys.95.045006}
}

@article{Panayi2014,
   author = {Christiana Panayi and Mohsen Razavi and Xiongfeng Ma and Norbert Lütkenhaus},
   doi = {10.1088/1367-2630/16/4/043005},
   issn = {13672630},
   journal = {New Journal of Physics},
   publisher = {IOP Publishing},
   title = {Memory-assisted measurement-device-independent quantum key distribution},
   volume = {16},
   year = {2014},
}

@article{Briegel1998,
   author = {H.-J. Briegel and W. Dür and J. I. Cirac and P. Zoller},
   doi = {10.1103/PhysRevLett.81.5932},
   issue = {26},
   journal = {Physical Review Letters},
   pages = {5932-5935},
   title = {Quantum Repeaters: The Role of Imperfect Local Operations in Quantum Communication},
   volume = {81},
   url = {http://link.aps.org/doi/10.1103/PhysRevLett.81.5932},
   year = {1998},
}

@article{Pompili2021,
  title={Realization of a multinode quantum network of remote solid-state qubits},
  author={Pompili, Matteo and Hermans, Sophie LN and Baier, Simon and Beukers, Hans KC and Humphreys, Peter C and Schouten, Raymond N and Vermeulen, Raymond FL and Tiggelman, Marijn J and dos Santos Martins, Laura and Dirkse, Bas and others},
  journal={Science},
  volume={372},
  number={6539},
  pages={259--264},
  year={2021},
  publisher={American Association for the Advancement of Science}
}

@article{Liu2021,
   author = {Xiao Liu and Jun Hu and Zong Feng Li and Xue Li and Pei Yun Li and Peng Jun Liang and Zong Quan Zhou and Chuan Feng Li and Guang Can Guo},
   doi = {10.1038/s41586-021-03505-3},
   issn = {14764687},
   issue = {7861},
   journal = {Nature},
   pages = {41-45},
   pmid = {34079139},
   publisher = {Springer US},
   title = {Heralded entanglement distribution between two absorptive quantum memories},
   volume = {594},
   url = {http://dx.doi.org/10.1038/s41586-021-03505-3},
   year = {2021},
}

@article{Lago2021,
   author = {Dario Lago-Rivera and Samuele Grandi and Jelena V. Rakonjac and Alessandro Seri and Hugues de Riedmatten},
   doi = {10.1038/s41586-021-03481-8},
   issn = {14764687},
   issue = {7861},
   journal = {Nature},
   pages = {37-40},
   pmid = {34079135},
   publisher = {Springer US},
   title = {Telecom-heralded entanglement between multimode solid-state quantum memories},
   volume = {594},
   url = {http://dx.doi.org/10.1038/s41586-021-03481-8},
   year = {2021},
}

@article{lu2026DI-QKD,
  title={Device-independent quantum key distribution over 100 km with single atoms},
  author={Lu, Bo-Wei and Yang, Chao-Wei and Wang, Run-Qi and Gao, Bo-Feng and Zhen, Yi-Zheng and Wang, Zhen-Gang and Shi, Jia-Kai and Ren, Zhong-Qi and Hahn, Thomas A and Tan, Ernest Y-Z and others},
  journal={Science},
  volume={391},
  number={6785},
  pages={592--597},
  year={2026},
  publisher={American Association for the Advancement of Science}
}

@article{liu2026long,
  title={Long-lived remote ion-ion entanglement for scalable quantum repeaters},
  author={Liu, Wen-Zhao and Zhou, Ya-Bin and Chen, Jiu-Peng and Wang, Bin and Teng, Ao and Han, Xiao-Wen and Liu, Guang-Cheng and Zhang, Zhi-Jiong and Yang, Yi and Liu, Feng-Guang and others},
  journal={Nature},
  pages={1--3},
  year={2026},
  publisher={Nature Publishing Group UK London}
}

@article{Vasconcelos2020,
   author = {Rui Vasconcelos and Sarah Reisenbauer and Cameron Salter and Georg Wachter and Daniel Wirtitsch and Jörg Schmiedmayer and Philip Walther and Michael Trupke},
   doi = {10.1038/s41534-019-0236-x},
   issn = {20566387},
   issue = {1},
   journal = {npj Quantum Information},
   publisher = {Nature Research},
   title = {Scalable spin–photon entanglement by time-to-polarization conversion},
   volume = {6},
   year = {2020},
}

@article{Lukin2020,
  title={Integrated quantum photonics with silicon carbide: challenges and prospects},
  author={Lukin, Daniil M and Guidry, Melissa A and Vu{\v{c}}kovi{\'c}, Jelena},
  journal={PRX Quantum},
  volume={1},
  number={2},
  pages={020102},
  year={2020},
  publisher={APS},
}

@article{Castelletto2022,
  title={Silicon carbide photonics bridging quantum technology},
  author={Castelletto, Stefania and Peruzzo, Alberto and Bonato, Cristian and Johnson, Brett C and Radulaski, Marina and Ou, Haiyan and Kaiser, Florian and Wrachtrup, Joerg},
  journal={ACS Photonics},
  volume={9},
  number={5},
  pages={1434--1457},
  year={2022},
  publisher={ACS Publications}
}

@article{Chatzopoulos2019,
  title={High-Q/V photonic crystal cavities and QED analysis in 3C-SiC},
  author={Chatzopoulos, Ioannis and Martini, Francesco and Cernansky, Robert and Politi, Alberto},
  journal={ACS Photonics},
  volume={6},
  number={8},
  pages={1826--1831},
  year={2019},
  publisher={ACS Publications}
}

@article{Faraon2013,
  title={Quantum photonic devices in single-crystal diamond},
  author={Faraon, Andrei and Santori, Charles and Huang, Zhihong and Fu, Kai-Mei C and Acosta, Victor M and Fattal, David and Beausoleil, Raymond G},
  journal={New Journal of Physics},
  volume={15},
  number={2},
  pages={025010},
  year={2013},
  publisher={IOP Publishing}
}

@article{Reiserer2015,
  title = {Cavity-based quantum networks with single atoms and optical photons},
  author = {Reiserer, Andreas and Rempe, Gerhard},
  journal = {Rev. Mod. Phys.},
  volume = {87},
  issue = {4},
  pages = {1379--1418},
  numpages = {40},
  year = {2015},
  month = {Dec},
  publisher = {American Physical Society},
  doi = {10.1103/RevModPhys.87.1379},
  url = {https://link.aps.org/doi/10.1103/RevModPhys.87.1379}
}

@article{Merkel2020,
  title={Coherent and Purcell-enhanced emission from erbium dopants in a cryogenic high-Q resonator},
  author={Merkel, Benjamin and Ulanowski, Alexander and Reiserer, Andreas},
  journal={Physical Review X},
  volume={10},
  number={4},
  pages={041025},
  year={2020},
  publisher={APS}
}

@article{Song2019,
  title={Ultrahigh-Q photonic crystal nanocavities based on 4H silicon carbide},
  author={Song, Bong-Shik and Asano, Takashi and Jeon, Seungwoo and Kim, Heungjoon and Chen, Changxuan and Kang, Dongyeon Daniel and Noda, Susumu},
  journal={Optica},
  volume={6},
  number={8},
  pages={991--995},
  year={2019},
  publisher={Optica Publishing Group}
}

@article{Fait2021,
    author = {Fait, J. and Putz, S. and Wachter, G. and Schalko, J. and Schmid, U. and Arndt, M. and Trupke, M.},
    title = "{High finesse microcavities in the optical telecom O-band}",
    journal = {Applied Physics Letters},
    volume = {119},
    number = {22},
    pages = {221112},
    year = {2021},
    month = {12},
    issn = {0003-6951},
    doi = {10.1063/5.0066620},
    url = {https://doi.org/10.1063/5.0066620},
}

@article{Derntl2014,
   author = {Christian Derntl and Michael Schneider and Johannes Schalko and Achim Bittner and Jörg Schmiedmayer and Ulrich Schmid and Michael Trupke},
   doi = {10.1364/oe.22.022111},
   issn = {10944087},
   issue = {18},
   journal = {Optics Express},
   pages = {22111},
   publisher = {The Optical Society},
   title = {Arrays of open, independently tunable microcavities},
   volume = {22},
   year = {2014},
}

@article{Bhaskar2020,
  doi = {10.1038/s41586-020-2103-5},
  url = {https://doi.org/10.1038/s41586-020-2103-5},
  year = {2020},
  publisher = {Springer Science and Business Media {LLC}},
  volume = {580},
  number = {7801},
  pages = {60--64},
  author = {M. K. Bhaskar and R. Riedinger and B. Machielse and D. S. Levonian and C. T. Nguyen and E. N. Knall and H. Park and D. Englund and M. Lon{\v{c}}ar and D. D. Sukachev and M. D. Lukin},
  title = {Experimental demonstration of memory-enhanced quantum communication},
  journal = {Nature}
}

@article{Spindlberger2019,
  doi = {10.1103/physrevapplied.12.014015},
  url = {https://doi.org/10.1103/physrevapplied.12.014015},
  year = {2019},
  publisher = {American Physical Society ({APS})},
  volume = {12},
  number = {1},
  author = {L. Spindlberger and A. Cs{\'{o}}r{\'{e}} and G. Thiering and S. Putz and R. Karhu and J.Ul Hassan and N.T. Son and T. Fromherz and A. Gali and M. Trupke},
  title = {Optical Properties of Vanadium in 4 H Silicon Carbide for Quantum Technology},
  journal = {Physical Review Applied}
}

@article{Mu2020,
	title = {Coherent {Manipulation} with {Resonant} {Excitation} and {Single} {Emitter} {Creation} of {Nitrogen} {Vacancy} {Centers} in {4H} {Silicon} {Carbide}},
	volume = {20},
	issn = {1530-6984},
	url = {https://doi.org/10.1021/acs.nanolett.0c02342},
	doi = {10.1021/acs.nanolett.0c02342},
	number = {8},
	urldate = {2021-06-05},
	journal = {Nano Letters},
	author = {Mu, Zhao and Zargaleh, Soroush Abbasi and von Bardeleben, Hans J\"urgen and Fr\"och, Johannes E. and Nonahal, Milad and Cai, Hongbing and Yang, Xinge and Yang, Jianqun and Li, Xingji and Aharonovich, Igor and Gao, Weibo},
	month = aug,
	year = {2020},
	pages = {6142--6147},
}

@article{Csore2017,
	title = {Characterization and formation of {NV} centers in ${3C}, {4H}$, and ${6H}$ {SiC}: {An} ab initio study},
	volume = {96},
	url = {https://link.aps.org/doi/10.1103/PhysRevB.96.085204},
	doi = {10.1103/PhysRevB.96.085204},
	number = {8},
	urldate = {2019-08-23},
	journal = {Physical Review B},
	author = {Cs\'or\'e, A. and von Bardeleben, H. J. and Cantin, J. L. and Gali, A.},
	year = {2017},
	pages = {085204},
}

@article{Bardeleben2016,
	title = {{NV} centers in ${3C},{4H}$, and ${6H}$ silicon carbide: {A} variable platform for solid-state qubits and nanosensors},
	volume = {94},
	shorttitle = {{NV} centers in \${3C},{4H}\$, and \${6H}\$ silicon carbide},
	url = {https://link.aps.org/doi/10.1103/PhysRevB.94.121202},
	doi = {10.1103/PhysRevB.94.121202},
	number = {12},
	urldate = {2019-08-23},
	journal = {Physical Review B},
	author = {von Bardeleben, H. J. and Cantin, J. L. and Cs\'or\'e, A. and Gali, A. and Rauls, E. and Gerstmann, U.},
	month = sep,
	year = {2016},
	pages = {121202},
}

@article{Gordon2015,
	title = {Defects as qubits in 3${C}$- and 4${H}$-{SiC}},
	volume = {92},
	url = {https://link.aps.org/doi/10.1103/PhysRevB.92.045208},
	doi = {10.1103/PhysRevB.92.045208},
	number = {4},
	urldate = {2019-08-23},
	journal = {Physical Review B},
	author = {Gordon, L. and Janotti, A. and Van de Walle, C. G.},
	month = jul,
	year = {2015},
	pages = {045208},
}

@article{WangPRL2020,
	title = {Coherent {Control} of {Nitrogen}-{Vacancy} {Center} {Spins} in {Silicon} {Carbide} at {Room} {Temperature}},
	volume = {124},
	url = {https://link.aps.org/doi/10.1103/PhysRevLett.124.223601},
	doi = {10.1103/PhysRevLett.124.223601},
	number = {22},
	urldate = {2020-06-11},
	journal = {Physical Review Letters},
	author = {Wang, Jun-Feng and Yan, Fei-Fei and Li, Qiang and Liu, Zheng-Hao and Liu, He and Guo, Guo-Ping and Guo, Li-Ping and Zhou, Xiong and Cui, Jin-Ming and Wang, Jian and Zhou, Zong-Quan and Xu, Xiao-Ye and Xu, Jin-Shi and Li, Chuan-Feng and Guo, Guang-Can},
	month = jun,
	year = {2020},
	pages = {223601},
}

@article{WangACSPhotonics2020,
	title = {Experimental {Optical} {Properties} of {Single} {Nitrogen} {Vacancy} {Centers} in {Silicon} {Carbide} at {Room} {Temperature}},
	volume = {7},
	url = {https://doi.org/10.1021/acsphotonics.0c00218},
	doi = {10.1021/acsphotonics.0c00218},
	number = {7},
	urldate = {2020-08-06},
	journal = {ACS Photonics},
	author = {Wang, Jun-Feng and Liu, Zheng-Hao and Yan, Fei-Fei and Li, Qiang and Yang, Xin-Ge and Guo, Liping and Zhou, Xiong and Huang, Wei and Xu, Jin-Shi and Li, Chuan-Feng and Guo, Guang-Can},
	month = jul,
	year = {2020},
	pages = {1611--1616},
}

@article{Murzakhanov2021,
	title = {Hyperfine and nuclear quadrupole splitting of the {NV}${^{-}}$ ground state in ${4H}$-{SiC}},
	volume = {103},
	url = {https://link.aps.org/doi/10.1103/PhysRevB.103.245203},
	doi = {10.1103/PhysRevB.103.245203},
	number = {24},
	urldate = {2021-06-24},
	journal = {Physical Review B},
	author = {Murzakhanov, F. F. and Yavkin, B. V. and Mamin, G. V. and Orlinskii, S. B. and von Bardeleben, H. J. and Biktagirov, T. and Gerstmann, U. and Soltamov, V. A.},
	month = jun,
	year = {2021},
	pages = {245203},
}

@article{Bardeleben2021,
	title = {Spin {Polarization}, {Electron}--{Phonon} {Coupling}, and {Zero}-{Phonon} {Line} of the {NV} {Center} in {3C}-{SiC}},
	volume = {21},
	issn = {1530-6984},
	url = {https://doi.org/10.1021/acs.nanolett.1c02564},
	doi = {10.1021/acs.nanolett.1c02564},
	number = {19},
	urldate = {2022-01-10},
	journal = {Nano Letters},
	author = {Jurgen von Bardeleben, Hans and Cantin, Jean-Louis and Gerstmann, Uwe and Schmidt, Wolf Gero and Biktagirov, Timur},
	year = {2021},
	pages = {8119--8125},
}

@article{Narahara2021,
	title = {Influences of hydrogen ion irradiation on N$_{\text{C}}$V$_{\text{Si}}^{-}$ formation in {4H}-silicon carbide},
	volume = {14},
	issn = {1882-0786},
	url = {https://dx.doi.org/10.35848/1882-0786/abdc9e},
	doi = {10.35848/1882-0786/abdc9e},
	OPTabstract = {Nitrogen-vacancy (NCVSi) center in 4H-SiC is spin defect with near-infrared luminescence at room temperature and a promising candidate for quantum technologies. This paper reports on NCVSi â’ center formation in N-doped 4H-SiCs by hydrogen ion irradiation and subsequent thermal annealing. It is revealed photoluminescence for NCVSi’ centers suddenly appears above the fluence of 5.0 Ă— 1015 cmâ’2 when annealed at 1000 Â°C. Appearance of a threshold fluence for their formation and/or activation has not been observed for other energetic particle irradiations. The possible mechanism is discussed based on the kinetics of hydrogen-related complexes and the majority carrier depletion caused by irradiation induced damage.},
	
	number = {2},
	urldate = {2023-08-27},
	journal = {Applied Physics Express},
	author = {Narahara, Takuma and Sato, Shin-ichiro and Kojima, Kazutoshi and Hijikata, Yasuto and Ohshima, Takeshi},
	year = {2021},
	pages = {021004},
	OPTfile = {IOP Full Text PDF:C\:\\Users\\gali.adam\\Zotero\\storage\\YXLT9UNW\\Narahara Ă©s mtsai. - 2021 - Influences of hydrogen ion irradiation on NcVsiâ’ f.pdf:application/pdf},
}

@article{Kaufmann1995,
	title = {Zeeman {Spectroscopy} and {Crystal}-{Field} {Model} of {Neutral} {Vandium} in {6H}-{Silicon} {Carbide}},
	url = {https://www.scientific.net/MSF.196-201.707},
	
	urldate = {2019-08-25},
	journal = {Materials Science Forum},
	author = {Kaufmann, B. and D\"ornen, Achim and Ham, Frank S.},
	year = {1995},
	volume = {196--201},
	pages = {707--712},
	doi = {10.4028/www.scientific.net/MSF.196-201.707},
}

@article{Baur1997,
	title = {Transition {Metals} in {SiC} {Polytypes}, as {Studied} by {Magnetic} {Resonance} {Techniques}},
	volume = {162},
	issn = {1521-396X},
	url = {https://onlinelibrary.wiley.com/doi/abs/10.1002/1521-396X%28199707%29162%3A1%3C153%3A%3AAID-PSSA153%3E3.0.CO%3B2-3},
	doi = {10.1002/1521-396X(199707)162:1<153::AID-PSSA153>3.0.CO;2-3},
	number = {1},
	urldate = {2019-08-25},
	journal = {physica status solidi (a)},
	author = {Baur, J. and Kunzer, M. and Schneider, J.},
	year = {1997},
	pages = {153--172},
}

@article{Schneider1990,
	title = {Infrared spectra and electron spin resonance of vanadium deep level impurities in silicon carbide},
	volume = {56},
	issn = {0003-6951},
	url = {https://aip.scitation.org/doi/10.1063/1.102555},
	doi = {10.1063/1.102555},
	number = {12},
	urldate = {2019-08-25},
	journal = {Applied Physics Letters},
	author = {Schneider, J. and M\"uller, H. D. and Maier, K. and Wilkening, W. and Fuchs, F. and D\"ornen, A. and Leibenzeder, S. and Stein, R.},
	month = mar,
	year = {1990},
	pages = {1184--1186},
}

@article{Wolfowicz2020,
	title = {Vanadium spin qubits as telecom quantum emitters in silicon carbide},
	volume = {6},
	copyright = {Copyright Â© 2020 The Authors, some rights reserved; exclusive licensee American Association for the Advancement of Science. No claim to original U.S. Government Works. Distributed under a Creative Commons Attribution NonCommercial License 4.0 (CC BY-NC).. This is an open-access article distributed under the terms of the Creative Commons Attribution-NonCommercial license, which permits use, distribution, and reproduction in any medium, so long as the resultant use is not for commercial advantage and provided the original work is properly cited.},
	issn = {2375-2548},
	url = {https://advances.sciencemag.org/content/6/18/eaaz1192},
	doi = {10.1126/sciadv.aaz1192},
	
	number = {18},
	urldate = {2020-08-11},
	journal = {Science Advances},
	author = {Wolfowicz, Gary and Anderson, Christopher P. and Diler, Berk and Poluektov, Oleg G. and Heremans, F. Joseph and Awschalom, David D.},
	month = may,
	year = {2020},
	pages = {eaaz1192},
}

@article{Csore2020,
	title = {Ab initio determination of pseudospin for paramagnetic defects in {SiC}},
	volume = {102},
	url = {https://link.aps.org/doi/10.1103/PhysRevB.102.241201},
	doi = {10.1103/PhysRevB.102.241201},
	number = {24},
	urldate = {2020-12-08},
	journal = {Physical Review B},
	author = {Cs\'or\'e, Andr\'as and Gali, Adam},
	year = {2020},
	pages = {241201},
}

@article{Tissot2021,
  title = {Hyperfine structure of transition metal defects in SiC},
  author = {Tissot, Benedikt and Burkard, Guido},
  journal = {Phys. Rev. B},
  volume = {104},
  issue = {6},
  pages = {064102},
  numpages = {10},
  year = {2021},
  month = {Aug},
  publisher = {American Physical Society},
  doi = {10.1103/PhysRevB.104.064102},
  url = {https://link.aps.org/doi/10.1103/PhysRevB.104.064102}
}

@article{Tissot2022,
	title = {Nuclear spin quantum memory in silicon carbide},
	volume = {4},
	url = {https://link.aps.org/doi/10.1103/PhysRevResearch.4.033107},
	doi = {10.1103/PhysRevResearch.4.033107},
	number = {3},
	urldate = {2023-08-27},
	journal = {Physical Review Research},
	author = {Tissot, Benedikt and Trupke, Michael and Koller, Philipp and Astner, Thomas and Burkard, Guido},
	month = aug,
	year = {2022},
	pages = {033107},
}

@article{Cilibrizzi2023,
   author = {Pasquale Cilibrizzi and Muhammad Junaid Arshad and Benedikt Tissot and Nguyen Tien Son and Ivan G. Ivanov and Thomas Astner and Philipp Koller and Misagh Ghezellou and Jawad Ul-Hassan and Daniel White and Christiaan Bekker and Guido Burkard and Michael Trupke and Cristian Bonato},
   doi = {10.1038/s41467-023-43923-7},
   issn = {2041-1723},
   issue = {1},
   journal = {Nature Communications},
   month = {12},
   pages = {8448},
   title = {Ultra-narrow inhomogeneous spectral distribution of telecom-wavelength vanadium centres in isotopically-enriched silicon carbide},
   volume = {14},
   year = {2023},
}

@article{bekker_SILs_2023,
    author = {Bekker, Christiaan and Arshad, Muhammad Junaid and Cilibrizzi, Pasquale and Nikolatos, Charalampos and Lomax, Peter and Wood, Graham S. and Cheung, Rebecca and Knolle, Wolfgang and Ross, Neil and Gerardot, Brian and Bonato, Cristian},
    title = "{Scalable fabrication of hemispherical solid immersion lenses in silicon carbide through grayscale hard-mask lithography}",
    journal = {Applied Physics Letters},
    volume = {122},
    number = {17},
    pages = {173507},
    year = {2023},
    month = {04},
    issn = {0003-6951},
    doi = {10.1063/5.0144684}
}

@misc{Hendriks2022,
	title = {Coherent spin dynamics of hyperfine-coupled vanadium impurities in silicon carbide},
	url = {http://arxiv.org/abs/2210.09942},
	doi = {10.48550/arXiv.2210.09942},
	urldate = {2023-08-27},
	publisher = {arXiv},
	author = {Hendriks, Joop and Gilardoni, Carmem M. and Adambukulam, Chris and Laucht, Arne and van der Wal, Caspar H.},
	month = oct,
	year = {2022},
    archivePrefix = {arXiv},
    eprint = {2210.09942},
}

@misc{Astner2022,
	title = {Vanadium in {Silicon} {Carbide}: {Telecom}-ready spin centres with long relaxation lifetimes and hyperfine-resolved optical transitions},
	shorttitle = {Vanadium in {Silicon} {Carbide}},
	url = {http://arxiv.org/abs/2206.06240},
	doi = {10.48550/arXiv.2206.06240},
	urldate = {2023-08-27},
	publisher = {arXiv},
	author = {Astner, T. and Koller, P. and Gilardoni, C. M. and Hendriks, J. and Son, N. T. and Ivanov, I. G. and Hassan, J. U. and van der Wal, C. H. and Trupke, M.},
	month = jun,
	year = {2022},
    archivePrefix = {arXiv},
    eprint = {2206.06240}
}

@article{Gilardoni2021,
	title = {Hyperfine-mediated transitions between electronic spin-1/2 levels of transition metal defects in {SiC}},
	volume = {23},
	issn = {1367-2630},
	url = {https://dx.doi.org/10.1088/1367-2630/ac1641},
	doi = {10.1088/1367-2630/ac1641},
	
	number = {8},
	urldate = {2023-08-27},
	journal = {New Journal of Physics},
	author = {Gilardoni, Carmem M. and Ion, Irina and Hendriks, Freddie and Trupke, Michael and Wal, Caspar H. van der},
	year = {2021},
	pages = {083010},
}

@article{Falk2015,
	title = {Optical {Polarization} of {Nuclear} {Spins} in {Silicon} {Carbide}},
	volume = {114},
	url = {http://link.aps.org/doi/10.1103/PhysRevLett.114.247603},
	doi = {10.1103/PhysRevLett.114.247603},
	number = {24},
	urldate = {2016-03-04},
	journal = {Physical Review Letters},
	author = {Falk, Abram L. and Klimov, Paul V. and Iv\'ady, Viktor and Sz\'asz, Kriszti\'an and Christle, David J. and Koehl, {William F.} and Gali, \'Ad\'am and Awschalom, David D.},
	month = jun,
	year = {2015},
	pages = {247603},
}

@article{Christle2015,
	title = {Isolated electron spins in silicon carbide with millisecond coherence times},
	volume = {14},
	copyright = {Â© 2014 Nature Publishing Group},
	issn = {1476-1122},
	url = {http://www.nature.com/nmat/journal/v14/n2/full/nmat4144.html},
	doi = {10.1038/nmat4144},
	
	number = {2},
	urldate = {2016-07-11},
	journal = {Nature Materials},
	author = {Christle, David J. and Falk, Abram L. and Andrich, Paolo and Klimov, Paul V. and Hassan, Jawad Ul and Son, Nguyen T. and Janz\'en, Erik and Ohshima, Takeshi and Awschalom, David D.},
	year = {2015},
	pages = {160--163},
}

@article{Gali2011,
	title = {Time-dependent density functional study on the excitation spectrum of point defects in semiconductors},
	volume = {248},
	copyright = {Copyright Â© 2011 WILEY-VCH Verlag GmbH \& Co. KGaA, Weinheim},
	issn = {1521-3951},
	url = {http://onlinelibrary.wiley.com/doi/10.1002/pssb.201046254/abstract},
	doi = {10.1002/pssb.201046254},
	
	number = {6},
	urldate = {2016-03-04},
	journal = {physica status solidi (b)},
	author = {Gali, Adam},
	month = jun,
	year = {2011},
	pages = {1337--1346},
}

@article{Christle2017,
	title = {Isolated {Spin} {Qubits} in {SiC} with a {High}-{Fidelity} {Infrared} {Spin}-to-{Photon} {Interface}},
	volume = {7},
	url = {https://link.aps.org/doi/10.1103/PhysRevX.7.021046},
	doi = {10.1103/PhysRevX.7.021046},
	number = {2},
	urldate = {2018-09-18},
	journal = {Physical Review X},
	author = {Christle, David J. and Klimov, Paul V. and de las Casas, Charles F. and Sz\'asz, Kriszti\'an and Iv\'ady, Viktor and Jokubavicius, Valdas and Ul Hassan, Jawad and Syv\"aj\"arvi, Mikael and Koehl, William F. and Ohshima, Takeshi and Son, Nguyen T. and Janz\'en, Erik and Gali, \'Ad\'am and Awschalom, David D.},
	month = jun,
	year = {2017},
	pages = {021046},
}

@article{Gali2010,
	title = {Theory of {Neutral} {Divacancy} in {SiC}: {A} {Defect} for {Spintronics}},
	volume = {645-648},
	issn = {1662-9752},
	shorttitle = {Theory of {Neutral} {Divacancy} in {SiC}},
	url = {http://www.scientific.net/MSF.645-648.395},
	doi = {10.4028/www.scientific.net/MSF.645-648.395},
	urldate = {2016-03-04},
	journal = {Materials Science Forum},
	author = {Gali, Adam and G\"allstr\"om, Andreas and Son, Nguyen T. and Janz\'en, Erik},
	month = apr,
	year = {2010},
	pages = {395--397},
}

@article{Koehl2011,
	title = {Room temperature coherent control of defect spin qubits in silicon carbide},
	volume = {479},
	copyright = {Â© 2011 Nature Publishing Group, a division of Macmillan Publishers Limited. All Rights Reserved.},
	issn = {0028-0836},
	url = {http://www.nature.com/nature/journal/v479/n7371/full/nature10562.html},
	doi = {10.1038/nature10562},
	
	number = {7371},
	urldate = {2016-03-04},
	journal = {Nature},
	author = {Koehl, William F. and Buckley, Bob B. and Heremans, F. Joseph and Calusine, Greg and Awschalom, David D.},
	year = {2011},
	pages = {84--87},
}

@article{Ivady2016,
	title = {High-{Fidelity} {Bidirectional} {Nuclear} {Qubit} {Initialization} in {SiC}},
	volume = {117},
	url = {https://link.aps.org/doi/10.1103/PhysRevLett.117.220503},
	doi = {10.1103/PhysRevLett.117.220503},
	number = {22},
	urldate = {2019-08-23},
	journal = {Physical Review Letters},
	author = {Iv\'ady, Viktor and Klimov, Paul V. and Miao, Kevin C. and Falk, Abram L. and Christle, David J. and Sz\'asz, Kriszti\'an and Abrikosov, Igor A. and Awschalom, David D. and Gali, Adam},
	year = {2016},
	pages = {220503},
}

@article{Ivady2019,
	title = {Stabilization of point-defect spin qubits by quantum wells},
	volume = {10},
	copyright = {2019 The Author(s)},
	issn = {2041-1723},
	url = {https://www.nature.com/articles/s41467-019-13495-6},
	doi = {10.1038/s41467-019-13495-6},
	
	number = {1},
	urldate = {2020-05-11},
	journal = {Nature Communications},
	author = {Iv\'ady, Viktor and Davidsson, Joel and Delegan, Nazar and Falk, Abram L. and Klimov, Paul V. and Whiteley, Samuel J. and Hruszkewycz, Stephan O. and Holt, Martin V. and Heremans, F. Joseph and Son, Nguyen Tien and Awschalom, David D. and Abrikosov, Igor A. and Gali, Adam},
	month = dec,
	year = {2019},
	pages = {5607},
}

@article{Crook2020,
	title = {Purcell {Enhancement} of a {Single} {Silicon} {Carbide} {Color} {Center} with {Coherent} {Spin} {Control}},
	volume = {20},
	issn = {1530-6984},
	url = {https://doi.org/10.1021/acs.nanolett.0c00339},
	doi = {10.1021/acs.nanolett.0c00339},
	number = {5},
	urldate = {2020-08-06},
	journal = {Nano Letters},
	author = {Crook, Alexander L. and Anderson, Christopher P. and Miao, Kevin C. and Bourassa, Alexandre and Lee, Hope and Bayliss, Sam L. and Bracher, David O. and Zhang, Xingyu and Abe, Hiroshi and Ohshima, Takeshi and Hu, Evelyn L. and Awschalom, David D.},
	month = may,
	year = {2020},
	pages = {3427--3434},
}

@article{Janzen2001,
	series = {Advanced {Characterisation} of {Semiconductor} {Materials}},
	title = {Material characterization need for {SiC}-based devices},
	volume = {4},
	issn = {1369-8001},
	url = {https://www.sciencedirect.com/science/article/pii/S1369800100001359},
	doi = {10.1016/S1369-8001(00)00135-9},
	
	number = {1},
	urldate = {2021-11-28},
	journal = {Materials Science in Semiconductor Processing},
	author = {Janz\'en, E. and Henry, A. and Bergman, J. P. and Ellison, A. and Magnusson, B.},
	year = {2001},
	pages = {181--186},
}

@article{Li2022,
  title={Room-temperature coherent manipulation of single-spin qubits in silicon carbide with a high readout contrast},
  author={Li, Qiang and Wang, Jun-Feng and Yan, Fei-Fei and Zhou, Ji-Yang and Wang, Han-Feng and Liu, He and Guo, Li-Ping and Zhou, Xiong and Gali, Adam and Liu, Zheng-Hao and others},
  journal={National Science Review},
  volume={9},
  number={5},
  pages={nwab122},
  year={2022},
  publisher={Oxford University Press}
}

@article{Anderson2019,
	title = {Electrical and optical control of single spins integrated in scalable semiconductor devices},
	volume = {366},
	url = {https://www.science.org/doi/10.1126/science.aax9406},
	doi = {10.1126/science.aax9406},
	number = {6470},
	urldate = {2022-02-23},
	journal = {Science},
	author = {Anderson, Christopher P. and Bourassa, Alexandre and Miao, Kevin C. and Wolfowicz, Gary and Mintun, Peter J. and Crook, Alexander L. and Abe, Hiroshi and Ul Hassan, Jawad and Son, Nguyen T. and Ohshima, Takeshi and Awschalom, David D.},
	year = {2019},
	pages = {1225--1230},
}

@article{Bourassa2020,
	title = {Entanglement and control of single nuclear spins in isotopically engineered silicon carbide},
	volume = {19},
	copyright = {2020 The Author(s), under exclusive licence to Springer Nature Limited},
	issn = {1476-4660},
	url = {https://www.nature.com/articles/s41563-020-00802-6},
	doi = {10.1038/s41563-020-00802-6},
	
	number = {12},
	urldate = {2022-11-20},
	journal = {Nature Materials},
	author = {Bourassa, Alexandre and Anderson, Christopher P. and Miao, Kevin C. and Onizhuk, Mykyta and Ma, He and Crook, Alexander L. and Abe, Hiroshi and Ul-Hassan, Jawad and Ohshima, Takeshi and Son, Nguyen T. and Galli, Giulia and Awschalom, David D.},
	month = dec,
	year = {2020},
	pages = {1319--1325},
}

@article{Son2006,
	title = {Divacancy in {4H}-{SiC}},
	volume = {96},
	url = {https://link.aps.org/doi/10.1103/PhysRevLett.96.055501},
	doi = {10.1103/PhysRevLett.96.055501},
	number = {5},
	urldate = {2023-08-27},
	journal = {Physical Review Letters},
	author = {Son, N. T. and Carlsson, P. and ul Hassan, J. and Janz\'en, E. and Umeda, T. and Isoya, J. and Gali, A. and Bockstedte, M. and Morishita, N. and Ohshima, T. and Itoh, H.},
	month = feb,
	year = {2006},
	pages = {055501},
}

@article{Widmann2015,
	title = {Coherent control of single spins in silicon carbide at room temperature},
	volume = {14},
	copyright = {Â© 2014 Nature Publishing Group},
	issn = {1476-1122},
	url = {http://www.nature.com/nmat/journal/v14/n2/abs/nmat4145.html},
	doi = {10.1038/nmat4145},
	
	number = {2},
	urldate = {2016-03-04},
	journal = {Nature Materials},
	author = {Widmann, Matthias and Lee, Sang-Yun and Rendler, Torsten and Son, Nguyen Tien and Fedder, Helmut and Paik, Seoyoung and Yang, Li-Ping and Zhao, Nan and Yang, Sen and Booker, Ian and Denisenko, Andrej and Jamali, Mohammad and Momenzadeh, S. Ali and Gerhardt, Ilja and Ohshima, Takeshi and Gali, Adam and Janz\'en, Erik and Wrachtrup, J\"org},
	year = {2015},
	pages = {164--168},
}

@article{widmann2019,
  title={Electrical charge state manipulation of single silicon vacancies in a silicon carbide quantum optoelectronic device},
  author={Widmann, Matthias and Niethammer, Matthias and Fedyanin, Dmitry Yu and Khramtsov, Igor A and Rendler, Torsten and Booker, Ian D and Ul Hassan, Jawad and Morioka, Naoya and Chen, Yu-Chen and Ivanov, Ivan G and others},
  journal={Nano letters},
  volume={19},
  number={10},
  pages={7173--7180},
  year={2019},
  publisher={ACS Publications}
}

@article{Gali2012,
	title = {Excitation spectrum of point defects in semiconductors studied by time-dependent density functional theory},
	volume = {27},
	issn = {2044-5326},
	url = {http://journals.cambridge.org/article_S0884291411004316},
	doi = {10.1557/jmr.2011.431},
	number = {06},
	urldate = {2016-03-04},
	journal = {Journal of Materials Research},
	author = {Gali, Adam},
	year = {2012},
	pages = {897--909},
}

@article{Ivady2017,
	title = {Identification of {Si}-vacancy related room-temperature qubits in ${4H}$ silicon carbide},
	volume = {96},
	url = {https://link.aps.org/doi/10.1103/PhysRevB.96.161114},
	doi = {10.1103/PhysRevB.96.161114},
	number = {16},
	urldate = {2018-09-26},
	journal = {Physical Review B},
	author = {Iv\'ady, Viktor and Davidsson, Joel and Son, Nguyen Tien and Ohshima, Takeshi and Abrikosov, Igor A. and Gali, Adam},
	month = oct,
	year = {2017},
	pages = {161114},
}

@article{Janzen2009,
	title = {The silicon vacancy in {SiC}},
	volume = {404},
	issn = {09214526},
	url = {http://linkinghub.elsevier.com/retrieve/pii/S0921452609010862},
	doi = {10.1016/j.physb.2009.09.023},
	
	number = {22},
	urldate = {2018-12-14},
	journal = {Physica B: Condensed Matter},
	author = {Janz\'en, Erik and Gali, Adam and Carlsson, Patrick and G\"allstr\"om, Andreas and Magnusson, Bj\"orn and Son, N.T.},
	month = dec,
	year = {2009},
	pages = {4354--4358},
}

@article{Dong2019,
	title = {Spin polarization through intersystem crossing in the silicon vacancy of silicon carbide},
	volume = {99},
	url = {https://link.aps.org/doi/10.1103/PhysRevB.99.184102},
	doi = {10.1103/PhysRevB.99.184102},
	number = {18},
	urldate = {2019-08-23},
	journal = {Physical Review B},
	author = {Dong, Wenzheng and Doherty, M. W. and Economou, Sophia E.},
	month = may,
	year = {2019},
	pages = {184102},
}

@article{Riedel2012,
	title = {Resonant {Addressing} and {Manipulation} of {Silicon} {Vacancy} {Qubits} in {Silicon} {Carbide}},
	volume = {109},
	url = {https://link.aps.org/doi/10.1103/PhysRevLett.109.226402},
	doi = {10.1103/PhysRevLett.109.226402},
	number = {22},
	urldate = {2019-08-26},
	journal = {Physical Review Letters},
	author = {Riedel, D. and Fuchs, F. and Kraus, H. and VĂ¤th, S. and Sperlich, A. and Dyakonov, V. and Soltamova, A. A. and Baranov, P. G. and Ilyin, V. A. and Astakhov, G. V.},
	year = {2012},
	pages = {226402},
}

@article{Baranov2011,
	title = {Silicon vacancy in {SiC} as a promising quantum system for single-defect and single-photon spectroscopy},
	volume = {83},
	url = {https://link.aps.org/doi/10.1103/PhysRevB.83.125203},
	doi = {10.1103/PhysRevB.83.125203},
	number = {12},
	urldate = {2019-08-26},
	journal = {Physical Review B},
	author = {Baranov, Pavel G. and Bundakova, Anna P. and Soltamova, Alexandra A. and Orlinskii, Sergei B. and Borovykh, Igor V. and Zondervan, Rob and Verberk, Rogier and Schmidt, Jan},
	month = mar,
	year = {2011},
	pages = {125203},
}

@article{Bracher2015,
   author = {David O. Bracher and Evelyn L. Hu},
   doi = {10.1021/acs.nanolett.5b02542},
   issn = {1530-6984},
   issue = {9},
   journal = {Nano Letters},
   month = {9},
   pages = {6202-6207},
   title = {Fabrication of High- <i>Q</i> Nanobeam Photonic Crystals in Epitaxially Grown 4H-SiC},
   volume = {15},
   year = {2015},
}

@article{mokhtarzadeh2022,
  title={Optimization of etching processes for the fabrication of smooth silicon carbide membranes for applications in quantum technology},
  author={Mokhtarzadeh, Mahsa and Carulla, Maria and Kozak, Roksolana and David, Christian},
  journal={Micro and Nano Engineering},
  volume={16},
  pages={100155},
  year={2022},
  publisher={Elsevier}
}

@article{Radulaski2017,
	title = {Scalable {Quantum} {Photonics} with {Single} {Color} {Centers} in {Silicon} {Carbide}},
	volume = {17},
	issn = {1530-6984},
	url = {https://doi.org/10.1021/acs.nanolett.6b05102},
	doi = {10.1021/acs.nanolett.6b05102},
	number = {3},
	urldate = {2019-08-28},
	journal = {Nano Letters},
	author = {Marina Radulaski and Matthias Widmann and Matthias Niethammer and Jingyuan Linda Zhang and Sang-Yun Lee and Torsten Rendler and Konstantinos G. Lagoudakis and Nguyen Tien Son and Erik Janz{\'{e}}n and Takeshi Ohshima and J\"{o}rg Wrachtrup and Jelena Vu{\v{c}}kovi{\'{c}}},
	month = mar,
	year = {2017},
	pages = {1782--1786},
}

@article{Soltamov2012,
	title = {Room {Temperature} {Coherent} {Spin} {Alignment} of {Silicon} {Vacancies} in ${4H}$- and ${6H}$-{SiC}},
	volume = {108},
	url = {https://link.aps.org/doi/10.1103/PhysRevLett.108.226402},
	doi = {10.1103/PhysRevLett.108.226402},
	number = {22},
	urldate = {2019-08-29},
	journal = {Physical Review Letters},
	author = {Soltamov, Victor A. and Soltamova, Alexandra A. and Baranov, Pavel G. and Proskuryakov, Ivan I.},
	month = may,
	year = {2012},
	pages = {226402},
}

@article{Soykal2016,
	title = {Silicon vacancy center in ${4H}$-{SiC}: {Electronic} structure and spin-photon interfaces},
	volume = {93},
	shorttitle = {Silicon vacancy center in \${4H}\$-{SiC}},
	url = {https://link.aps.org/doi/10.1103/PhysRevB.93.081207},
	doi = {10.1103/PhysRevB.93.081207},
	number = {8},
	urldate = {2019-08-29},
	journal = {Physical Review B},
	author = {Soykal, \"O. O. and Dev, Pratibha and Economou, Sophia E.},
	year = {2016},
	pages = {081207},
}

@article{Soykal2017,
	title = {Quantum metrology with a single spin-$\frac{3}{2}$ defect in silicon carbide},
	volume = {95},
	url = {https://link.aps.org/doi/10.1103/PhysRevB.95.081405},
	doi = {10.1103/PhysRevB.95.081405},
	number = {8},
	urldate = {2019-08-29},
	journal = {Physical Review B},
	author = {Soykal, \"O. O. and Reinecke, T. L.},
	year = {2017},
	pages = {081405},
}

@article{Nagy2019,
	title = {High-fidelity spin and optical control of single silicon-vacancy centres in silicon carbide},
	volume = {10},
	copyright = {2019 The Author(s)},
	issn = {2041-1723},
	url = {https://www.nature.com/articles/s41467-019-09873-9},
	doi = {10.1038/s41467-019-09873-9},
	
	number = {1},
	urldate = {2019-08-30},
	journal = {Nature Communications},
	author = {Nagy, Roland and Niethammer, Matthias and Widmann, Matthias and Chen, Yu-Chen and Udvarhelyi, P\'eter and Bonato, Cristian and Hassan, Jawad Ul and Karhu, Robin and Ivanov, Ivan G. and Son, Nguyen Tien and Maze, Jeronimo R. and Ohshima, Takeshi and Soykal, \"Oney O. and Gali, \'Ad\'am and Lee, Sang-Yun and Kaiser, Florian and Wrachtrup, J\"org},
	month = apr,
	year = {2019},
	pages = {1954},
}

@article{Wang2021,
	title = {Robust coherent control of solid-state spin qubits using anti-{Stokes} excitation},
	volume = {12},
	copyright = {2021 The Author(s)},
	issn = {2041-1723},
	url = {https://www.nature.com/articles/s41467-021-23471-8},
	doi = {10.1038/s41467-021-23471-8},
	number = {1},
	urldate = {2021-11-29},
	journal = {Nature Communications},
	author = {Wang, Jun-Feng and Yan, Fei-Fei and Li, Qiang and Liu, Zheng-Hao and Cui, Jin-Ming and Liu, Zhao-Di and Gali, Adam and Xu, Jin-Shi and Li, Chuan-Feng and Guo, Guang-Can},
	month = may,
	year = {2021},
	pages = {3223},
}

@article{Babin2022,
	title = {Fabrication and nanophotonic waveguide integration of silicon carbide colour centres with preserved spin-optical coherence},
	volume = {21},
	copyright = {2021 The Author(s), under exclusive licence to Springer Nature Limited},
	issn = {1476-4660},
	url = {https://www.nature.com/articles/s41563-021-01148-3},
	doi = {10.1038/s41563-021-01148-3},
	number = {1},
	urldate = {2022-02-23},
	journal = {Nature Materials},
	author = {Babin, Charles and St\"ohr, Rainer and Morioka, Naoya and Linkewitz, Tobias and Steidl, Timo and W\"ornle, Raphael and Liu, Di and Hesselmeier, Erik and Vorobyov, Vadim and Denisenko, Andrej and Hentschel, Mario and Gobert, Christian and Berwian, Patrick and Astakhov, Georgy V. and Knolle, Wolfgang and Majety, Sridhar and Saha, Pranta and Radulaski, Marina and Son, Nguyen Tien and Ul-Hassan, Jawad and Kaiser, Florian and Wrachtrup, J\"org},
	month = jan,
	year = {2022},
	pages = {67--73},
}

@article{Ruf2021,
	title = {Quantum networks based on color centers in diamond},
	volume = {130},
	issn = {0021-8979},
	url = {https://aip.scitation.org/doi/10.1063/5.0056534},
	doi = {10.1063/5.0056534},
	number = {7},
	urldate = {2021-08-30},
	journal = {Journal of Applied Physics},
	author = {Ruf, Maximilian and Wan, Noel H. and Choi, Hyeongrak and Englund, Dirk and Hanson, Ronald},
	year = {2021},
	pages = {070901},
}

@article{Becker2016,
	title = {Ultrafast all-optical coherent control of single silicon vacancy colour centres in diamond},
	volume = {7},
	url = {http://dx.doi.org/10.1038/ncomms13512},
	journal = {Nature Communications},
	author = {Becker, Jonas Nils and G\"orlitz, Johannes and Arend, Carsten and Markham, Matthew and Becher, Christoph},
	year = {2016},
	pages = {13512},
}

@article{Thiering2018,
	title = {Ab {Initio} {Magneto}-{Optical} {Spectrum} of {Group}-{IV} {Vacancy} {Color} {Centers} in {Diamond}},
	volume = {8},
	url = {https://link.aps.org/doi/10.1103/PhysRevX.8.021063},
	doi = {10.1103/PhysRevX.8.021063},
	number = {2},
	urldate = {2019-09-09},
	journal = {Physical Review X},
	author = {Thiering, Gerg\H{o} and Gali, Adam},
	month = jun,
	year = {2018},
	pages = {021063},
}

@article{Goss1996,
	title = {The {Twelve}-{Line} 1.682 {eV} {Luminescence} {Center} in {Diamond} and the {Vacancy}-{Silicon} {Complex}},
	volume = {77},
	url = {https://link.aps.org/doi/10.1103/PhysRevLett.77.3041},
	doi = {10.1103/PhysRevLett.77.3041},
	number = {14},
	urldate = {2019-09-09},
	journal = {Physical Review Letters},
	author = {Goss, J. P. and Jones, R. and Breuer, S. J. and Briddon, P. R. and Öberg, S.},
	month = sep,
	year = {1996},
	pages = {3041--3044},
}

@article{Gali2013,
	title = {Ab initio study of the split silicon-vacancy defect in diamond: {Electronic} structure and related properties},
	volume = {88},
	shorttitle = {Ab initio study of the split silicon-vacancy defect in diamond},
	url = {https://link.aps.org/doi/10.1103/PhysRevB.88.235205},
	doi = {10.1103/PhysRevB.88.235205},
	number = {23},
	urldate = {2019-09-09},
	journal = {Physical Review B},
	author = {Gali, Adam and Maze, Jeronimo R.},
	year = {2013},
	pages = {235205},
}

@article{Neu2011,
	title = {Single photon emission from silicon-vacancy colour centres in chemical vapour deposition nano-diamonds on iridium},
	volume = {13},
	issn = {1367-2630},
	url = {https://doi.org/10.1088%2F1367-2630%2F13%2F2%2F025012},
	doi = {10.1088/1367-2630/13/2/025012},
	number = {2},
	urldate = {2019-09-09},
	journal = {New Journal of Physics},
	author = {Neu, Elke and Steinmetz, David and Riedrich-M\"uller, Janine and Gsell, Stefan and Fischer, Martin and Schreck, Matthias and Becher, Christoph},
	year = {2011},
	pages = {025012},
}

@article{Hepp2014,
	title = {Electronic {Structure} of the {Silicon} {Vacancy} {Color} {Center} in {Diamond}},
	volume = {112},
	url = {https://link.aps.org/doi/10.1103/PhysRevLett.112.036405},
	doi = {10.1103/PhysRevLett.112.036405},
	number = {3},
	urldate = {2019-09-09},
	journal = {Physical Review Letters},
	author = {Hepp, Christian and M\"uller, Tina and Waselowski, Victor and Becker, Jonas N. and Pingault, Benjamin and Sternschulte, Hadwig and Steinm\"uller-Nethl, Doris and Gali, Adam and Maze, Jeronimo R. and Atat\"ure, Mete and Becher, Christoph},
	year = {2014},
	pages = {036405},
}

@article{Rogers2014,
	title = {All-{Optical} {Initialization}, {Readout}, and {Coherent} {Preparation} of {Single} {Silicon}-{Vacancy} {Spins} in {Diamond}},
	volume = {113},
	url = {https://link.aps.org/doi/10.1103/PhysRevLett.113.263602},
	doi = {10.1103/PhysRevLett.113.263602},
	number = {26},
	urldate = {2019-09-09},
	journal = {Physical Review Letters},
	author = {Rogers, Lachlan J. and Jahnke, Kay D. and Metsch, Mathias H. and Sipahigil, Alp and Binder, Jan M. and Teraji, Tokuyuki and Sumiya, Hitoshi and Isoya, Junichi and Lukin, Mikhail D. and Hemmer, Philip and Jelezko, Fedor},
	year = {2014},
	pages = {263602},
}

@article{Pingault2017,
	title = {Coherent control of the silicon-vacancy spin in diamond},
	volume = {8},
	copyright = {2017 Nature Publishing Group},
	issn = {2041-1723},
	url = {https://www.nature.com/articles/ncomms15579},
	doi = {10.1038/ncomms15579},
	
	urldate = {2019-09-09},
	journal = {Nature Communications},
	author = {Pingault, Benjamin and Jarausch, David-Dominik and Hepp, Christian and Klintberg, Lina and Becker, Jonas N. and Markham, Matthew and Becher, Christoph and Atat\"ure, Mete},
	month = may,
	year = {2017},
	pages = {15579},
}

@article{Sukachev2017,
	title = {Silicon-{Vacancy} {Spin} {Qubit} in {Diamond}: {A} {Quantum} {Memory} {Exceeding} 10 ms with {Single}-{Shot} {State} {Readout}},
	volume = {119},
	shorttitle = {Silicon-{Vacancy} {Spin} {Qubit} in {Diamond}},
	url = {https://link.aps.org/doi/10.1103/PhysRevLett.119.223602},
	doi = {10.1103/PhysRevLett.119.223602},
	number = {22},
	urldate = {2019-09-09},
	journal = {Physical Review Letters},
	author = {Sukachev, D. D. and Sipahigil, A. and Nguyen, C. T. and Bhaskar, M. K. and Evans, R. E. and Jelezko, F. and Lukin, M. D.},
	year = {2017},
	pages = {223602},
}

@article{Becker2018,
	title = {All-{Optical} {Control} of the {Silicon}-{Vacancy} {Spin} in {Diamond} at {Millikelvin} {Temperatures}},
	volume = {120},
	url = {https://link.aps.org/doi/10.1103/PhysRevLett.120.053603},
	doi = {10.1103/PhysRevLett.120.053603},
	number = {5},
	urldate = {2019-09-09},
	journal = {Physical Review Letters},
	author = {Becker, Jonas N. and Pingault, Benjamin and Gro{\ss}, David and G\"undo{\c g}an, Mustafa and Kukharchyk, Nadezhda and Markham, Matthew and Edmonds, Andrew and Atat\"ure, Mete and Bushev, Pavel and Becher, Christoph},
	year = {2018},
	pages = {053603},
}

@article{Sohn2018,
	title = {Controlling the coherence of a diamond spin qubit through its strain environment},
	volume = {9},
	copyright = {2018 The Author(s)},
	issn = {2041-1723},
	url = {https://www.nature.com/articles/s41467-018-04340-3},
	doi = {10.1038/s41467-018-04340-3},
	
	number = {1},
	urldate = {2019-11-13},
	journal = {Nature Communications},
	author = {Sohn, Young-Ik and Meesala, Srujan and Pingault, Benjamin and Atikian, Haig A. and Holzgrafe, Jeffrey and G\"undo{\c g}an, Mustafa and Stavrakas, Camille and Stanley, Megan J. and Sipahigil, Alp and Choi, Joonhee and Zhang, Mian and Pacheco, Jose L. and Abraham, John and Bielejec, Edward and Lukin, Mikhail D. and Atat\"ure, Mete and Lon{\c c}ar, Marko},
	month = may,
	year = {2018},
	pages = {1--6},
}

@article{Sternschulte1994,
	title = {1.681-{eV} luminescence center in chemical-vapor-deposited homoepitaxial diamond films},
	volume = {50},
	url = {https://link.aps.org/doi/10.1103/PhysRevB.50.14554},
	doi = {10.1103/PhysRevB.50.14554},
	number = {19},
	urldate = {2019-11-21},
	journal = {Physical Review B},
	author = {Sternschulte, H. and Thonke, K. and Sauer, R. and Münzinger, P. C. and Michler, P.},
	year = {1994},
	pages = {14554--14560},
}

@article{Machielse2019,
	title = {Quantum {Interference} of {Electromechanically} {Stabilized} {Emitters} in {Nanophotonic} {Devices}},
	volume = {9},
	url = {https://link.aps.org/doi/10.1103/PhysRevX.9.031022},
	doi = {10.1103/PhysRevX.9.031022},
	number = {3},
	urldate = {2019-11-22},
	journal = {Physical Review X},
	author = {Machielse, B. and Bogdanovic, S. and Meesala, S. and Gauthier, S. and Burek, M. J. and Joe, G. and Chalupnik, M. and Sohn, Y. I. and Holzgrafe, J. and Evans, R. E. and Chia, C. and Atikian, H. and Bhaskar, M. K. and Sukachev, D. D. and Shao, L. and Maity, S. and Lukin, M. D. and Loncar, M.},
	year = {2019},
	pages = {031022},
}

@article{Bersin2023,
  title = {Telecom Networking with a Diamond Quantum Memory},
  author = {Bersin, Eric and Sutula, Madison and Huan, Yan Qi and Suleymanzade, Aziza and Assumpcao, Daniel R. and Wei, Yan-Cheng and Stas, Pieter-Jan and Knaut, Can M. and Knall, Erik N. and Langrock, Carsten and Sinclair, Neil and Murphy, Ryan and Riedinger, Ralf and Yeh, Matthew and Xin, C.J. and Bandyopadhyay, Saumil and Sukachev, Denis D. and Machielse, Bartholomeus and Levonian, David S. and Bhaskar, Mihir K. and Hamilton, Scott and Park, Hongkun and Lon\ifmmode \check{c}\else \v{c}\fi{}ar, Marko and Fejer, Martin M. and Dixon, P. Benjamin and Englund, Dirk R. and Lukin, Mikhail D.},
  journal = {PRX Quantum},
  volume = {5},
  issue = {1},
  pages = {010303},
  numpages = {10},
  year = {2024},
  month = {Jan},
  publisher = {American Physical Society},
  doi = {10.1103/PRXQuantum.5.010303},
  url = {https://link.aps.org/doi/10.1103/PRXQuantum.5.010303}
}

@article{Pingault2014,
	title = {All-{Optical} {Formation} of {Coherent} {Dark} {States} of {Silicon}-{Vacancy} {Spins} in {Diamond}},
	volume = {113},
	url = {https://link.aps.org/doi/10.1103/PhysRevLett.113.263601},
	doi = {10.1103/PhysRevLett.113.263601},
	number = {26},
	urldate = {2023-08-27},
	journal = {Physical Review Letters},
	author = {Pingault, Benjamin and Becker, Jonas N. and Schulte, Carsten H. H. and Arend, Carsten and Hepp, Christian and Godde, Tillmann and Tartakovskii, Alexander I. and Markham, Matthew and Becher, Christoph and Atatüure, Mete},
	month = dec,
	year = {2014},
	pages = {263601},
}

@article{Wang2013midIR,
author = {Wang, Shunchong and Zhan, Minjie and Wang, Gang and Xuan, Hongwen and Zhang, Wei and Liu, Chunjun and Xu, Chunhua and Liu, Yu and Wei, Zhiyi and Chen, Xiaolong},
title = {4H-SiC: a new nonlinear material for midinfrared lasers},
journal = {Laser Photonics Rev.},
volume = {7},
number = {5},
pages = {831-838},
year = {2013}
}

@article{zheng20194h,
  title={4H-SiC microring resonators for nonlinear integrated photonics},
  author={Zheng, Yi and Pu, Minhao and Yi, Ailun and Ou, Xin and Ou, Haiyan},
  journal={Opt. Lett.},
  volume={44},
  number={23},
  pages={5784--5787},
  year={2019},
  publisher={Optical Society of America}
}

@article{zheng2019high,
  title={High-quality factor, high-confinement microring resonators in 4H-silicon carbide-on-insulator},
  author={Zheng, Yi and Pu, Minhao and Yi, Ailun and Chang, Bingdong and You, Tiangui and Huang, Kai and Kamel, Ayman N and Henriksen, Martin R and J{\o}rgensen, Asbj{\o}rn A and Ou, Xin and others},
  journal={Opt. Express},
  volume={27},
  number={9},
  pages={13053--13060},
  year={2019},
  publisher={Optical Society of America}
}

@article{sato2009accurate,
  title={Accurate measurements of second-order nonlinear optical coefficients of 6H and 4H silicon carbide},
  author={Sato, Hiroaki and Abe, Makoto and Shoji, Ichiro and Suda, Jun and Kondo, Takashi},
  journal={J. of the Opt. Society of America B},
  volume={26},
  number={10},
  pages={1892--1896},
  year={2009},
  publisher={Optical Society of America}
}

@article{Guidry_parametric_optica_2020,
author = {Melissa A. Guidry and Ki Youl Yang and Daniil M. Lukin and Ashot Markosyan and Joshua Yang and Martin M. Fejer and Jelena Vu\v{c}kovi\'{c}},
journal = {Optica},
pages = {1139--1142},
title = {Optical parametric oscillation in silicon carbide nanophotonics},
volume = {7},
year = {2020},
url = {http://www.osapublishing.org/optica/abstract.cfm?URI=optica-7-9-1139},
doi = {10.1364/OPTICA.394138}
}

@article{nagy_PrAppl2023,
  title = {Scalable Quantum Memory Nodes Using Nuclear Spins in Silicon Carbide},
  author = {Parthasarathy, Shravan Kumar and Kallinger, Birgit and Kaiser, Florian and Berwian, Patrick and Dasari, Durga B.R. and Friedrich, Jochen and Nagy, Roland},
  journal = {Phys. Rev. Appl.},
  volume = {19},
  issue = {3},
  pages = {034026},
  numpages = {8},
  year = {2023},
  month = {Mar},
  publisher = {American Physical Society},
  doi = {10.1103/PhysRevApplied.19.034026},
  url = {https://link.aps.org/doi/10.1103/PhysRevApplied.19.034026}
}

@article{cramer_repeated_2016,
	title = {Repeated quantum error correction on a continuously encoded qubit by real-time feedback},
	volume = {7},
	copyright = {2016 The Author(s)},
	issn = {2041-1723},
	url = {https://www.nature.com/articles/ncomms11526},
	doi = {10.1038/ncomms11526},
	
	number = {1},
	urldate = {2023-04-26},
	journal = {Nature Communications},
	author = {Cramer, J. and Kalb, N. and Rol, M. A. and Hensen, B. and Blok, M. S. and Markham, M. and Twitchen, D. J. and Hanson, R. and Taminiau, T. H.},
	month = may,
	year = {2016},
	pages = {11526},
}

@article{taminiau_PRX_10spins,
  title = {A Ten-Qubit Solid-State Spin Register with Quantum Memory up to One Minute},
  author = {Bradley, C. E. and Randall, J. and Abobeih, M. H. and Berrevoets, R. C. and Degen, M. J. and Bakker, M. A. and Markham, M. and Twitchen, D. J. and Taminiau, T. H.},
  journal = {Phys. Rev. X},
  volume = {9},
  issue = {3},
  pages = {031045},
  numpages = {12},
  year = {2019},
  month = {Sep},
  publisher = {American Physical Society},
  doi = {10.1103/PhysRevX.9.031045},
  url = {https://link.aps.org/doi/10.1103/PhysRevX.9.031045}
}

@article{High-fidelity_Read-out_2011,
	title = {High-fidelity projective read-out of a solid-state spin quantum register},
	volume = {477},
	copyright = {2011 Nature Publishing Group, a division of Macmillan Publishers Limited. All Rights Reserved.},
	issn = {1476-4687},
	url = {https://www.nature.com/articles/nature10401},
	doi = {10.1038/nature10401},
	
	number = {7366},
	urldate = {2023-01-26},
	journal = {Nature},
	author = {Robledo, Lucio and Childress, Lilian and Bernien, Hannes and Hensen, Bas and Alkemade, Paul F. A. and Hanson, Ronald},
	month = sep,
	year = {2011},
	pages = {574--578},
}

@article{Dreau_Hanson_2018,
  title = {Quantum Frequency Conversion of Single Photons from a Nitrogen-Vacancy Center in Diamond to Telecommunication Wavelengths},
  author = {Dr\'eau, Ana\"{\i}s and Tchebotareva, Anna and Mahdaoui, Aboubakr El and Bonato, Cristian and Hanson, Ronald},
  journal = {Phys. Rev. Appl.},
  volume = {9},
  issue = {6},
  pages = {064031},
  numpages = {8},
  year = {2018},
  month = {Jun},
  publisher = {American Physical Society},
  doi = {10.1103/PhysRevApplied.9.064031},
  url = {https://link.aps.org/doi/10.1103/PhysRevApplied.9.064031}
}

@article{Rakonjac_ICFO_2021,
  title = {Entanglement between a Telecom Photon and an On-Demand Multimode Solid-State Quantum Memory},
  author = {Rakonjac, Jelena V. and Lago-Rivera, Dario and Seri, Alessandro and Mazzera, Margherita and Grandi, Samuele and de Riedmatten, Hugues},
  journal = {Phys. Rev. Lett.},
  volume = {127},
  issue = {21},
  pages = {210502},
  numpages = {7},
  year = {2021},
  month = {Nov},
  publisher = {American Physical Society},
  doi = {10.1103/PhysRevLett.127.210502},
  url = {https://link.aps.org/doi/10.1103/PhysRevLett.127.210502}
}

@article{anderson_5s_SSRO_SiC2022,
	title = {Five-second coherence of a single spin with single-shot readout in silicon carbide},
	volume = {8},
	issn = {2375-2548},
	url = {https://www.science.org/doi/10.1126/sciadv.abm5912},
	doi = {10.1126/sciadv.abm5912},
	
	number = {5},
	urldate = {2023-09-06},
	journal = {Science Advances},
	author = {Anderson, Christopher P. and Glen, Elena O. and Zeledon, Cyrus and Bourassa, Alexandre and Jin, Yu and Zhu, Yizhi and Vorwerk, Christian and Crook, Alexander L. and Abe, Hiroshi and Ul-Hassan, Jawad and Ohshima, Takeshi and Son, Nguyen T. and Galli, Giulia and Awschalom, David D.},
	month = feb,
	year = {2022},
	pages = {eabm5912},
}

@article{rogge_Er_SSRO2013,
	title = {Optical addressing of an individual erbium ion in silicon},
	volume = {497},
	copyright = {2013 Springer Nature Limited},
	issn = {1476-4687},
	url = {https://www.nature.com/articles/nature12081},
	doi = {10.1038/nature12081},
	
	number = {7447},
	urldate = {2023-09-06},
	journal = {Nature},
	author = {Yin, Chunming and Rancic, Milos and de Boo, Gabriele G. and Stavrias, Nikolas and McCallum, Jeffrey C. and Sellars, Matthew J. and Rogge, Sven},
	month = may,
	year = {2013},
	pages = {91--94},
}

@article{zhang_SSRO2021,
	title = {High-fidelity single-shot readout of single electron spin in diamond with spin-to-charge conversion},
	volume = {12},
	copyright = {2021 The Author(s)},
	issn = {2041-1723},
	url = {https://www.nature.com/articles/s41467-021-21781-5},
	doi = {10.1038/s41467-021-21781-5},
	
	number = {1},
	urldate = {2023-09-06},
	journal = {Nature Communications},
	author = {Zhang, Qi and Guo, Yuhang and Ji, Wentao and Wang, Mengqi and Yin, Jun and Kong, Fei and Lin, Yiheng and Yin, Chunming and Shi, Fazhan and Wang, Ya and Du, Jiangfeng},
	month = mar,
	year = {2021},
	pages = {1529},
}

@article{wengerowsky2018entanglement,
  title={An entanglement-based wavelength-multiplexed quantum communication network},
  author={Wengerowsky, S{\"o}ren and Joshi, Siddarth Koduru and Steinlechner, Fabian and H{\"u}bel, Hannes and Ursin, Rupert},
  journal={Nature},
  volume={564},
  number={7735},
  pages={225--228},
  year={2018},
  publisher={Nature Publishing Group UK London}
}

@article{hall2021survey,
  title={A survey of reconfigurable optical networks},
  author={Hall, Matthew Nance and Foerster, Klaus-Tycho and Schmid, Stefan and Durairajan, Ramakrishnan},
  journal={Optical Switching and Networking},
  volume={41},
  pages={100621},
  year={2021},
  publisher={Elsevier}
}

@article{pirandola2017fundamental,
  title={Fundamental limits of repeaterless quantum communications},
  author={Pirandola, Stefano and Laurenza, Riccardo and Ottaviani, Carlo and Banchi, Leonardo},
  journal={Nature communications},
  volume={8},
  number={1},
  pages={15043},
  year={2017},
  publisher={Nature Publishing Group UK London}
}

@article{bader2024analysis,
  title={Analysis, recent challenges and capabilities of spin-photon interfaces in Silicon carbide-on-insulator},
  author={Bader, Joshua and Arianfard, Hamed and Peruzzo, Alberto and Castelletto, Stefania},
  journal={npj Nanophotonics},
  volume={1},
  number={1},
  pages={29},
  year={2024},
  publisher={Nature Publishing Group UK London}
}

@article{zhou2025SiCNetworks,
  title={Silicon carbide: A promising platform for scalable quantum networks},
  author={Zhou, Yu and Tan, Junhua and Hu, HaiBo and Hua, Sikai and Jiang, Chunhui and Liang, Bo and Bao, Tongyuan and Nie, Xinfang and Xiao, Shumin and Lu, Dawei and Wang, Junfeng and Song, Qinghai},
  journal={Applied Physics Reviews},
  volume={12},
  number={3},
  year={2025},
  publisher={AIP Publishing}
}

@article{he2026spinPhoton,
  title={Coherent Spin-Photon Interface of single PL6 Color Centers in Silicon Carbide},
  author={He, Zhen-Xuan and Thiering, Gerg{\H{o}} and Liang, Rui-Jian and Zhou, Ji-Yang and Ren, Shuo and Lin, Wu-Xi and Hao, Zhi-He and Hu, Qi-Cheng and Wang, Jun-Feng and Gali, Adam and others},
  journal={arXiv preprint arXiv:2602.06421},
  year={2026}
}

@misc{Fang2023,
      title={Experimental Generation of Spin-Photon Entanglement in Silicon Carbide}, 
      author={Ren-Zhou Fang and Xiao-Yi Lai and Tao Li and Ren-Zhu Su and Bo-Wei Lu and Chao-Wei Yang and Run-Ze Liu and Yu-Kun Qiao and Cheng Li and Zhi-Gang He and Jia Huang and Hao Li and Li-Xing You and Yong-Heng Huo and Xiao-Hui Bao and Jian-Wei Pan},
      year={2023},
      eprint={2311.17455},
      archivePrefix={arXiv},
      primaryClass={quant-ph}
}

@article{Anisimov2025chlorine,
  title={Engineering chlorine-based emitters in silicon carbide for telecom-band quantum technologies},
  author={Anisimov, Andrey N and Mathews, Ashin V and Mavridou, Kalliopi and Kentsch, Ulrich and Helm, Manfred and Astakhov, Georgy V},
  journal={Optics Express},
  volume={33},
  number={26},
  pages={54285--54294},
  year={2025},
  publisher={Optica Publishing Group}
}

@article{Shafizadeh2026vanadium3C,
  title={Vanadium photoluminescence in 3C-SiC},
  author={Shafizadeh, Danial and Jokubavicius, Valdas and Murata, Koichi and Tsuchida, Hidekazu and Udvarhelyi, P{\'e}ter and Bian, Guodong and Lang, Oliver and Karaman, Merve and Enriquez, Diego Haya and Brehm, Moritz and others},
  journal={Physical Review B},
  volume={113},
  number={7},
  pages={075201},
  year={2026},
  publisher={APS}
}

\newpage
\onecolumngrid
\appendix
\clearpage
\section{Parameter list for MA-MDI-QKD simulations}
\label{sec:app}
Here, we list all relevant parameters used for the simulations of the MA-MDI-QKD protocol using SiC devices. 
The values are derived from state-of-the-art experiments and off-the-shelf component specifications. 
The first part of Table \ref{tab:param} lists the spin-cavity parameters, while the second part of the table lists parameters used in the simulation. 

\begin{table}[h]
    \centering
    \caption{Parameters used in the MA-MDI-QKD simulation with SiC devices.}
    \label{tab:param}
    \rowcolors{1}{gray!30}{white} 
    \begin{tabularx}{1\textwidth} { 
  | >{\hsize=0.4\hsize\raggedright\arraybackslash}X 
  | >{\hsize=0.1\hsize\centering\arraybackslash}X 
  | >{\hsize=0.1\hsize\centering\arraybackslash}X 
  | >{\hsize=0.4\hsize\raggedright\arraybackslash}X|}
        \hline
        \textbf{Name} & \textbf{Symbol} & \textbf{Value} & \textbf{Comment/Description} \\
        \hline
        Amplitude decay time & $T_1$ & 30s & Amplitude decay time of the electron spin $\eta_r(t) = \eta_{r0} \mathrm{e}^{-t/T_1}$. \\
        Coherence time & $T_2$ & 10ms-10s & Dephasing time of electron spin. \\
        Nuclear spin amplitude decay time & $T_{\mathrm{1N}}$ & 30s & Amplitude decay time of the nuclear spin. \\
        Nuclear spin coherence time & $T_{\mathrm{2N}}$ & $>10s$ & Dephasing time of nuclear spin. \\
        Spin conversion time & $T_{\mathrm{E-N}}$ & \SI{10}{\micro\second} &Time needed for a conversion from electron spin to nuclear spin. \\
        Spin readout time & $\tau_\mathrm{R}$ & \SI{400}{\nano\second} & Time required to measure the electron spin. \\
        $\pi$-pulse time & $\tau_\pi$ &10-100 ns& $\pi$-pulse duration. \\
        Initialization time & $\tau_\mathrm{init}$& $\tau_\mathrm{R}+\tau_\mathrm{\pi}$ & Initialization time for the electron spin. \\
        Reflectivity $\ket{\uparrow}$ & $\eta_\uparrow$ & 1 & Reflectivity of the cavity with the spin in the $\uparrow$ state, SMF included. \\
        Reflectivity $\ket{\downarrow}$ & $\eta_\downarrow$ & 0 & Reflectivity of the cavity with the spin in the $\downarrow$ state, SMF included. \\
        Linewidth & $\gamma$ & 100 MHz & Optical Fourier-limited linewidth (FWHM), including cooperative broadening. \\ \hline \hline 
        Pulse duration & $\tau_\mathrm{p}$ & 11.2 ns & Optical pulse duration. \\
        Writing time & $\tau_\mathrm{w}$ & 20-200 ns & $2\left(\tau_\mathrm{p} + \tau_\pi\right)$. Time difference between the time that the beginning of the pulse arrives at the QM and the time that a successful/unsuccessful loading is declared. \\
        Reading time & $\tau_\mathrm{r}$ & 410-500 ns & 400 ns + $\tau_\pi$. Time difference between the time that the retrieval process is applied until the end of the pulse is out. \\
        Writing efficiency & $\eta_\mathrm{w}$ & 0.13 & Probability to store a qubit and herald success conditioned on having a single photon at the QMs input.\\
        Reading efficiency & $\eta_{r0}$ & 1 & Probability to retrieve a single photon out of the QM (right after loading t=0) conditioned on a successful loading in the past. \\
        Single-photon detection efficiency & $\eta_\mathrm{SPD}$ & 0.85 & Superconducting nanowire single photon detector \\
        Temporal resolution & $t_\mathrm{SPD}$ & 50 ps & Superconducting nanowire single photon detector, FWHM \\
        Dark count rate & $\gamma_\mathrm{dc}$ & 100 Hz & \\
        Background rate & $\gamma_\mathrm{bg}$ & 0 Hz & Photonic background rate per optical mode \\
        Channel length & $L$ & 0 – 700 km & $L_\mathrm{A(B)}$ is the channel length from Alice or Bob to the respective memory. $L = L_\mathrm{A} + L_\mathrm{B}$, where we assume equal channel length $L_\mathrm{A} = L_\mathrm{B}$.  \\
        Fibre attenuation & $\alpha_\mathrm{Ob}$ & 0.3 dB/km & At the telecom O-band \\
        Setup misalignment probability & $e_\mathrm{A(B)}$ & 0 & Probability of a qubit flip \\
        Error correction inefficiency & $f$ & 1.16 & Efficiency of the error correction code \\
        Insertion loss fibre optical circulator & $\eta_\mathrm{oc}$& 0.8 dB & \\ 
        Insertion loss fibre optical switch  & $\eta_\mathrm{os}$& 0.6 dB & \\ 
        Optical rise/fall time optical switch &$t_\mathrm{os}$ & 8 ns & \\
        Minimal pulse width optical switch & $\Delta t_\mathrm{os}$ & 90 ns & \\
        \hline
    \end{tabularx}
\end{table}

\end{document}